\documentclass[aps,pra,twocolumn,showpacs,preprintnumbers,amsmath,amssymb,superscriptaddress,10pt,nofootinbib]{revtex4-1}

\usepackage{amsmath,amssymb}
\usepackage{bm}
\usepackage[latin1]{inputenc} 
\usepackage{dcolumn}
\usepackage{graphicx}
\usepackage{SIunits}
\usepackage{ae}
\usepackage{color} 

\newcommand{\diff}[0]{\text{d}}

\newcommand{\re}{\mathrm{Re}\,}
\newcommand{\im}{\mathrm{Im}\,}

\newcommand{\vekh}[1]{\hat{{\boldsymbol{\mathbf{#1}}}}}

\newcommand{\be}{\begin{equation}}
\newcommand{\ee}{\end{equation}}
\newcommand{\ba}{\begin{align}}
\newcommand{\ea}{\end{align}}
\newcommand{\e}[1]{\text{e}^{#1}}

\begin{document}

\title{Fourier theory of linear gain media}

\author{Hans Olaf Hågenvik}
\author{Markus E. Malema}
\author{Johannes Skaar}
\affiliation{Department of Electronics and Telecommunications, Norwegian University of Science and Technology, NO-7491 Trondheim, Norway}
\email{johannes.skaar@ntnu.no}

\date{\today}

\begin{abstract}
The analysis of wave propagation in linear, passive media is usually done by considering a single real frequency (the monochromatic limit) and also often a single plane wave component (plane wave limit). For gain media, we demonstrate that these two limits generally do not commute; for example, one order may lead to a diverging field, while the other order leads to a finite field. Moreover, the plane wave limit may be dependent on whether it is realized with a finite support excitation or gaussian excitation, eventually of infinite widths. We consider wave propagation in gain media by a Fourier--Laplace integral in space and time, and demonstrate how the correct monochromatic limit or plane wave limit can be taken, by deforming the integration surface in complex frequency--complex wavenumber space. We also give the most general criterion for absolute instabilities. The general theory is applied in several cases, and is used to predict media with novel properties. In particular, we show the existence of isotropic media which in principle exhibit simultaneous refraction, meaning that they refract positively and negatively at the same time.
\end{abstract}

\maketitle

\section{Introduction}
Fourier theory makes it possible to consider single frequencies and plane wave components separately, in describing electromagnetic wave propagation in linear, passive media. This leads to huge simplification in analysis and interpretation, especially for dispersive (and/or spatially dispersive) media. Nevertheless, we must have in mind that real physics happens in the time--spatial domain, not in frequency--wavenumber space; the monochromatic and plane wave limits can never be realized in practice. The monochromatic limit is approached by turning on the excitation at some time $t=0$ \cite{brillouin}, and waiting a sufficiently long time until the transients have died out. The plane wave limit is approached by letting the width of the excitation be sufficiently large.

For active media (gain media), it is clearly of large interest to use the same Fourier theory, by decomposing the field into frequency components and/or plane waves. There is, however, a number of obstacles. The most obvious one is that active media are inherently nonlinear due to gain saturation \cite{saleh}. In practice, this can be dealt with by verifying that the magnitude of the solution is less than the threshold for gain saturation. If it is not, then the excitation must be reduced accordingly, or the solution must be rejected. If there are divergences associated with the linear solution, the solution must be rejected in any case.

Another problem is that the Fourier transform not necessarily exists. A remedy is to use the Laplace transform, decomposing the time-domain fields into exponentially increasing functions $\exp(-i\omega t)$ for $\im\omega>0$ (see Sec. \ref{sec:laplfour}). Once the solution has been found, it can often be continuated towards real frequencies, enabling simpler interpretation (Sec. \ref{sec:cont}). One may argue that the Fourier transform should be sufficient for the relevant situations, since diverging solutions must be rejected anyway. However, this strategy is dangerous, as imposing Fourier transform analysis may give the impression of false, stable solutions.

An extensively discussed problem in the context of active media, is the determination of the sign of the longitudinal wavenumber $k_z$. This problem is far from trivial, even e.g. in the context of total internal reflection from a weakly  amplifying medium \cite{koester66, *romanov72, *kogan72, *lebedev73, *kolokolov75, *callary76, *lukosz76, *cybulski77, *silverman83t, *silverman83, *kolokolov98, *fan03, *siegman10, kolokolov99, willis08, grepstad11}. More recently, the problem has been discussed in the context of the wave vector or refractive index of more advanced active media including active metamaterials \cite{chen05, *mackay159701, *ramakrishna059701, *chen159702, *chen059702, *ramakrishna_ol, *nazarov07, *perezmolina08, *nazarov08, skaar06, skaar06b, geddes07}. 

We are not going to focus on this problem here, as it now seems to be agreement that the sign of the longitudinal wavenumber must be determined by ensuring it is analytic in some upper half-plane of complex frequency, and such that $k_z\to+\omega/c$ for $\omega\to\infty$ \cite{skaar06, skaar06b, geddes07, premaratne, afanasev2013}. Here $\omega$ is the (possibly complex) frequency and $c$ the vacuum light velocity. However, we will take the analysis one important step further; by considering a double Fourier--Laplace transform with respect to space and time. Clearly, for realistic situations, the fields can neither have infinite durations nor infinite widths. In addition to turning the field excitations on at $t=0$, it turns out to be crucial to let them have finite widths, to see how the medium behaves in practice. Indeed, even though a particular medium does not show absolute instabilities for plane wave excitations, it can support absolute instabilities in the presence of other excitations. 

Once the general theory governing causal finite beam propagation has been discussed, it is of interest to consider the monochromatic limit and plane wave limit. A number of peculiar but interesting results arise. First of all, the monochromatic and plane wave limits do not commute in general. For very common situations with conventional gain media, one order leads to finite fields, while the other order leads to infinite fields. Second of all, the plane wave limit may depend on the way it is taken; if it is realized using a finite-support excitation or a gaussian excitation, eventually of infinite widths. Our analysis leads to a better understanding of the nontrivialities associated with earlier, monochromatic and plane-wave analyses of active media. It also can be used to predict new classes of active media, with novel responses. For example, we predict the presence of isotropic media which exhibit simultaneous refraction, i.e., both positive and negative refraction simultaneously. While this is a novel and surprising response, it may be argued that the required gain is unrealistically high, and makes both realization and time-domain simulations challenging, at least for the specific media proposed here.

Previously, Kolokolov \cite{kolokolov99} and Grepstad and Skaar \cite{grepstad11} have treated the problem of Fourier--Laplace transform analysis of active media. However, Kolokolov only considered the special case with weak or no dispersion. Dispersion has important consequences for the theory, as it turns out to fundamentally change the method of deformation in the complex frequency--wavenumber space. The dispersion, possibly engineered by metamaterials, may lead to new classes of active media, as shown by the different possible behaviors in frequency--wavenumber space. Grepstad and Skaar did not perform a complete analysis, since they did not consider the deformation in frequency--wavenumber space, including the monochromatic limit for finite beams.

The article is structured as follows. In Sec. \ref{sec:laplfour}, we state the problem and discuss the assumptions in detail, before analyzing the fields using the Laplace transform (in time) and Fourier transform (in space). In Sec. \ref{sec:cont} we discuss how we may approach real frequencies for media without absolute instabilities. This happens at the expense of deforming the integration path in the complex wavenumber ($k_x$) space. In Sec. \ref{sec:realfreqplanewave} we discuss the plane wave limit, and the interpretation of divergences and non-commutativity. The theory is applied to the understanding of existing media and novel media in Sec. \ref{sec:examples}. In particular, we show the presence of simultaneous refraction, before concluding in Sec. \ref{sec:concl}.

\section{Laplace and Fourier transform analysis}\label{sec:laplfour}
We restrict the analysis to linear, time-shift invariant, isotropic, homogeneous media without spatial dispersion. Moreover, we assume the following asymptotic behavior for the product of relative permittivity $\epsilon$ and relative permeability $\mu$, as $\omega\to\infty$ \cite{landau_lifshitz_edcm}: $\epsilon(\omega)\mu(\omega)=1+\mathcal O(\omega^{-2})$. Finally, we assume that the medium does not support superexponential instabilities \cite{nistad08}, meaning that any field solution should not grow faster with time than an exponential.

In the analysis we consider an infinite or semi-infinite medium. Considering infinite media helps us understand the electromagnetic response given solely by the medium's properties -- effects related to interactions with surrounding media have been ruled out. Of course, there are no infinite gain media in practice.  However, as long as the smallest distance from an observation point to the boundary of the medium is larger than $ct_{\max}$, where $t_{\max}$ is the maximum duration of the experiment, the size does not matter and we may as well assume it is infinite. To approach steady state (or the monochromatic limit) we will later require $t_{\max}$ to be large. Then we must have in mind that the dimensions of the gain medium must be accordingly large. 

We will assume that the medium is dark for $t\leq 0$. This assumption needs some clarification. To establish the active medium, an energy pump must be turned on before $t=0$. When the system does not support instabilities, we can imagine that the pump was turned on a long time before $t=0$, such that any transients have died out. If there are instabilities, however, any disturbance will blow up with time. We could assume that the pump is turned on slowly before $t=0$, sufficiently smooth such that no significant transients are generated as a result of the pump, but sufficiently fast such that the (small) transients do not grow too much before $t=0$. We do not consider the existence of such a trade-off further; we rather demand that any transients from the pump or from other perturbations or fluctuations in the system, must be included into the analysis. This is done by including them into the excitation of the system, to be defined below.

It is also in order to comment the linearity assumptions in some detail. The amplitude in any practical medium will be limited by nonlinear effects such as gain saturation. When we refer to ``diverging fields'', or ``instabilities'', it strictly means that the fields grow until they are limited by gain saturation. Clearly, in such cases the linear analysis is only accurate for a limited duration. In the absence of instabilities, the analysis is clearly accurate for all times, provided the excitations are sufficiently weak.

\begin{figure}
\centering \includegraphics[width=4.6cm]{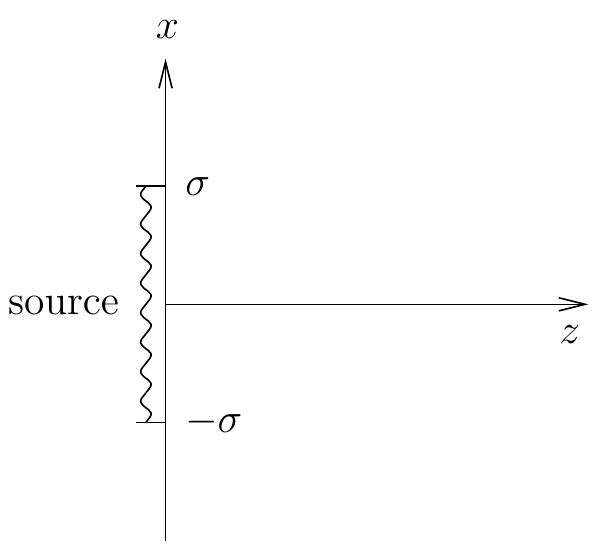}
\caption{An excitation is located at $z=0$ in a homogeneous medium. In the figure the special case with finite width $2\sigma$ is shown.}
\label{fig:setup}
\end{figure}
For simplicity we limit the discussion to propagation in two dimensions, $x$ and $z$, and transversal electric (TE) fields. Let $\mathcal E(x,z,t)\vekh y$ be the physical electric field, pointing in the $y$-direction $\vekh y$. Since the medium is active, the field may diverge with time $t$. We have limited our attention to active media and sources that lead to fields growing at most exponentially. Moreover, we assume that the electric field is square integrable (finite energy) with respect to $x$ (for the complete assumptions see Appendix \ref{sec:fubini2}). The electric field is Laplace transformable: 
\begin{equation}\label{lapl}
 E(x,z,\omega)=\int_0^\infty \mathcal E(x,z,t)\exp(i\omega t)\diff t,
\end{equation}
for $\im\omega>\gamma$, where $\gamma$ is a sufficiently large positive number characterizing the maximum growth of the field. Furthermore, $E(x,z,\omega)$ is Fourier transformed wrt. $x$, to obtain the plane wave spectrum
\begin{equation}\label{four}
 E(k_x,z,\omega)=\int_{-\infty}^\infty E(x,z,\omega)\exp(-ik_xx)\diff x.
\end{equation}

The inverse transform can be written
\begin{align}\label{invlaplfour}
& (2\pi)^2\mathcal E(x,z,t) = \\
& = \int_{i\gamma-\infty}^{i\gamma+\infty}\int_{-\infty}^\infty E(k_x,z,\omega)\exp(ik_xx-i\omega t)\diff k_x\diff\omega \nonumber\\
& = \int_{-\infty}^\infty\int_{i\gamma-\infty}^{i\gamma+\infty} E(k_x,z,\omega)\exp(ik_xx-i\omega t)\diff\omega\diff k_x, \nonumber
\end{align}
where, in the last equality, we have interchanged the order of integration (see Appendix \ref{sec:fubini2}). 

We consider a source in the plane $z=0$ (Fig. \ref{fig:setup}), infinitely thin, but possibly of infinite width. 
In general, we may have sources everywhere; in that case, we would have to superpose the fields resulting from the different sources. For $z\neq 0$, Maxwell's equations mean that $\left(\diff^2/\diff z^2-k_x^2+\epsilon\mu\omega^2/c^2\right)E(k_x,z,\omega)=0$. Furthermore, the transversal ($x$-component) of the magnetic field is given by $-i\omega\mu\mu_0 H(k_x,z,\omega)=\diff E(k_x,z,\omega)/\diff z$, where $\mu_0$ is the permeability in vacuum. Hence, we can express
\begin{subequations}\label{EH}
\begin{align}
E(k_x,z,\omega) &= A(k_x,\omega)\e{ik_zz}+B(k_x,\omega)\e{-ik_zz} \label{Ep}\\
H(k_x,z,\omega) &= -\frac{k_z}{\omega\mu\mu_0}\left[A(k_x,\omega)\e{ik_zz}-B(k_x,\omega)\e{-ik_zz}\right] \label{Hp}
\end{align}
\end{subequations}
for $z<0$, and
\begin{subequations}\label{EHm}
\begin{align}
E(k_x,z,\omega) &= C(k_x,\omega)\e{ik_zz}+D(k_x,\omega)\e{-ik_zz} \label{Em}\\
H(k_x,z,\omega) &= -\frac{k_z}{\omega\mu\mu_0}\left[C(k_x,\omega)\e{ik_zz}-D(k_x,\omega)\e{-ik_zz}\right] \label{Hm}
\end{align}
\end{subequations}
for $z>0$. Here
\begin{equation}\label{kz}
k_z^2=\epsilon\mu\frac{\omega^2}{c^2}-k_x^2.
\end{equation}
The four functions $A(k_x,\omega)$, $B(k_x,\omega)$, $C(k_x,\omega)$, and $D(k_x,\omega)$ are connected by the electromagnetic boundary conditions, which in turn, are dependent on the source. For a current source, $E(k_x,z,\omega)$ is continuous across the source plane, while $H(k_x,0^+,\omega)-H(k_x,0^-,\omega)=J(k_x,\omega)$, where $J(k_x,\omega)$ is the (Fourier--Laplace transformed) surface current source. With reflection symmetry about the plane $z=0$, this means that
\begin{subequations}\label{determABCD}
\begin{align}
A&=D, \\
B&=C, \\
\frac{2k_z}{\omega\mu\mu_0}(A-B)&=J(k_x,\omega). \label{determAB}
\end{align}
\end{subequations}
Clearly, both unknown functions $A$ and $B$ cannot be found from \eqref{determABCD}. Moreover, since the medium potentially is active, we cannot use principles like requiring the source to do positive work, or field decay as $z\to\infty$. We must invoke the principle of causality in its most fundamental form. 

First we note that the sign of $k_z$ can be chosen arbitrarily in \eqref{EH} and \eqref{EHm}; a change of sign means only that the functions $C$ and $D$ (and $A$ and $B$) are interchanged. Since $\epsilon(\omega)$ and $\mu(\omega)$ are analytic for $\im\omega>\gamma$, and tend to unity as $\omega\to\infty$ there, we choose the sign such that for a fixed $k_x$,
\begin{align}\label{kzsign}
& k_z(k_x,\omega) \text{ is analytic for } \im\omega > \gamma, \text{ and } \\
& k_z(k_x,\omega) \to+\omega/c \text{ as } \omega\to\infty \text{ in the region }\im\omega>\gamma. \nonumber
\end{align}
Assuming that the medium and the source are dark for $t<0$, the fields as described by \eqref{EHm} are causal, and we can use a version of Titchmarsh' theorem for diverging functions (Appendix \ref{sec:Titchdiv}) to prove that 
in \eqref{EH} and \eqref{EHm}, we have
\begin{subequations}\label{determABCDfin}
\begin{align}
A&=D=0, \label{determADfin}\\
B&=C=-\frac{\mu\mu_0\omega}{2k_z} J(k_x,\omega). \label{determBCfin}
\end{align}
\end{subequations}
Moreover, in Appendix \ref{sec:fubini} we prove that the function $k_z(k_x,\omega)$ is zero-free in a region $\im\omega > \gamma$; thus $B$ is analytic there\footnote{If we had chosen the opposite sign for $k_z$ in \eqref{kzsign}, we would have obtained $B=C=0$. If we had chosen the sign in another, arbitrary way, we would have obtained $A=D=0$ for some frequencies, and $B=C=0$ else. Such choices are inconvenient (but perfectly valid) as $k_z$ and the four functions $A$, $B$, $C$, and $D$ get nonanalytic.}.

We now consider the usual situation described by the Fresnel equations, where we have different media on each side of the plane $z=0$, and there is no source at $z=0$ but rather somewhere in the medium on the left-hand side ($z<0$). Clearly, we can use the identical causality argument on the right-hand side ($z>0$), to obtain \eqref{kzsign} and $D=0$. The electromagnetic boundary conditions $E(k_x,0^+,\omega)=E(k_x,0^-,\omega)$ and $H(k_x,0^+,\omega)=H(k_x,0^-,\omega)$ then give the reflection and transmission coefficients
\begin{subequations}\label{fresnel}
\begin{align}
\frac{B}{A} = \frac{\mu_2k_{1z}-\mu_1k_{2z}}{\mu_2k_{1z}+\mu_1k_{2z}},\\
\frac{C}{A}=\frac{2\mu_2k_{1z}}{\mu_2k_{1z}+\mu_1k_{2z}}, \label{fresnelC}
\end{align}
\end{subequations}
where $k_{iz}^2=\epsilon_i\mu_i\omega^2/c^2-k_x^2$. Here subscript 1 and 2 stand for the medium to the left and right, respectively. Throughout this paper, we will for simplicity assume that medium 1 is vacuum or a passive medium.

We will consider sources in the product form $u(x)v(t)$, with transform $U(k_x)V(\omega)$. For the situation with a current source plane, we set $J(k_x,\omega) = -U(k_x)V(\omega)/c\mu_0$, and for the situation with an incident wave, we set $A(k_x,\omega)=U(k_x)V(\omega)$. For later use, we sum up by writing the electric field solutions for $z>0$ for the current source plane and the Fresnel situation, respectively:
\begin{subequations}\label{Ezpos}
\begin{align}
E(k_x,z,\omega) &= \frac{\mu\omega\e{ik_zz}}{2k_zc}U(k_x)V(\omega), \label{Ezposa}\\ 
E(k_x,z,\omega) &= \frac{2\mu_2k_{1z}\e{ik_zz}}{\mu_2k_{1z}+\mu_1k_{2z}}U(k_x)V(\omega). \label{Ezposb}
\end{align}
\end{subequations}
Here $k_z$ is given by \eqref{kz} and \eqref{kzsign}. 
It is important to note that these results have been derived for $\im\omega>\gamma$. In Sec. \ref{sec:cont} we will consider the possibility of continuating the solutions towards real frequencies. 

\section{Towards real frequencies}\label{sec:cont}
To facilitate interpretation and computation, it is useful to examine if we can move the inverse Laplace transform contour (Bromwich path) in \eqref{invlaplfour} down to the real $\omega$-axis, such that it describes an inverse Fourier transform. This is desirable, as steady-state harmonic excitations and solutions are convenient to interpret physically. For the active media and systems where this is possible, we have only convective instabilities \cite{sturrock,briggs}: Then, nondiverging excitations lead to nondiverging fields for every fixed point $(x,z)$. This means that any growing wave must be convected away. On the other hand, if the Bromwich path cannot be moved down to the real axis due to singularities or cuts, the transform can be described as an inverse Fourier transform plus integrals around the nonanalytic points. Since the latter integrals diverge with time, we have absolute instabilities, meaning that the fields diverge even at fixed points in space.

For a wide range of active media of interest, it turns out to be possible to move the Bromwich path in \eqref{invlaplfour} down to the real axis, at the expense of deforming the integration path in the $k_x$-domain \cite{briggs}. This is what we will consider in the following. The clue here is to realize that the integrand is analytic in both $k_x$ and $\omega$, so integration paths can be deformed until they reach singularities. To this end we assume that $\epsilon\mu$ does not have singularities or zeros for $\im\omega\geq 0$; situations with zeros in the upper half-plane will be discussed later. Under these conditions $\sqrt{\epsilon\mu}$ is analytic and zero-free for $\im\omega\geq 0$. We consider the evaluation of the physical field in the spatial and time domain, according to \eqref{invlaplfour}, but along a possibly deformed surface $\Gamma$ in the $(k_x,\omega)$-domain:
\begin{equation}\label{invlaplfourGamma}
(2\pi)^2\mathcal E(x,z,t) = \int_{\Gamma} E(k_x,z,\omega)\exp(ik_xx-i\omega t)\diff k_x\diff\omega
\end{equation}
Here $E(k_x,z,\omega)$ is given by \eqref{Ezpos}. Apparently, the integrand is analytic in both $k_x$ and $\omega$, except at the branch cuts arising from the square root $k_z=\sqrt{\epsilon\mu\omega^2/c^2-k_x^2}$, and also if $k_z=0$ for the case \eqref{Ezposa}, or if $\mu_2k_{1z}+\mu_1k_{2z}=0$ for the case \eqref{Ezposb}. The last possibility will be ignored in the following; we simply assume that two involved media are chosen such that these singularities do not disturb the deformation of $\Gamma$. Examples will be given later. From the theory below it will also become clear how to generalize to account for such singularities.

\begin{figure}
\centering \includegraphics[width=8cm]{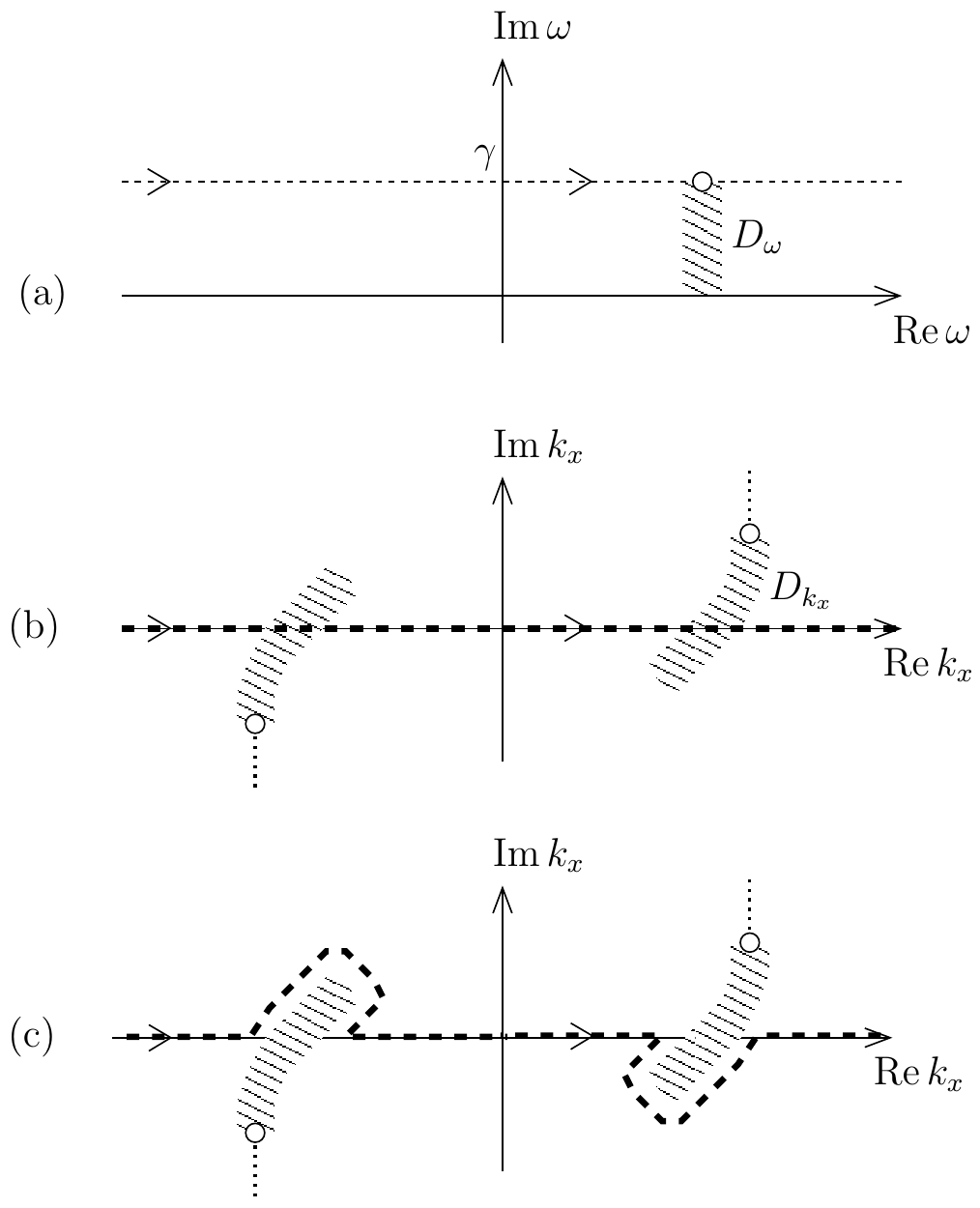}
\caption{The dashed lines correspond to the integration paths in \eqref{invlaplfour}: (a) $\omega$-domain; (b) $k_x$-domain; and (c) deformed path in the $k_x$-domain for the $\omega$ indicated by a circle in (a). The domain $D_{k_x}$ corresponds to the set of values $k_x=\pm\sqrt{\epsilon\mu}\omega/c$ for $\omega\in D_{\omega}$. The open circles in the $k_x$-plane correspond to the open circle in the $\omega$-plane. The dotted vertical lines indicate branch cuts for $k_z(k_x,\omega)$ for the particular $\omega$ as indicated by the open circle. We proved in Appendix \ref{sec:fubini} that $k_z(k_x,\omega)$ is analytic wrt. $k_x$, for $\im\omega=\gamma$ and real $k_x$; thus the branch cuts must avoid the real $k_x$-axis. In the figure we take them to be vertical, starting at the circles.} 
\label{fig:deformation}
\end{figure}

\begin{figure}
\centering \includegraphics[width=8cm]{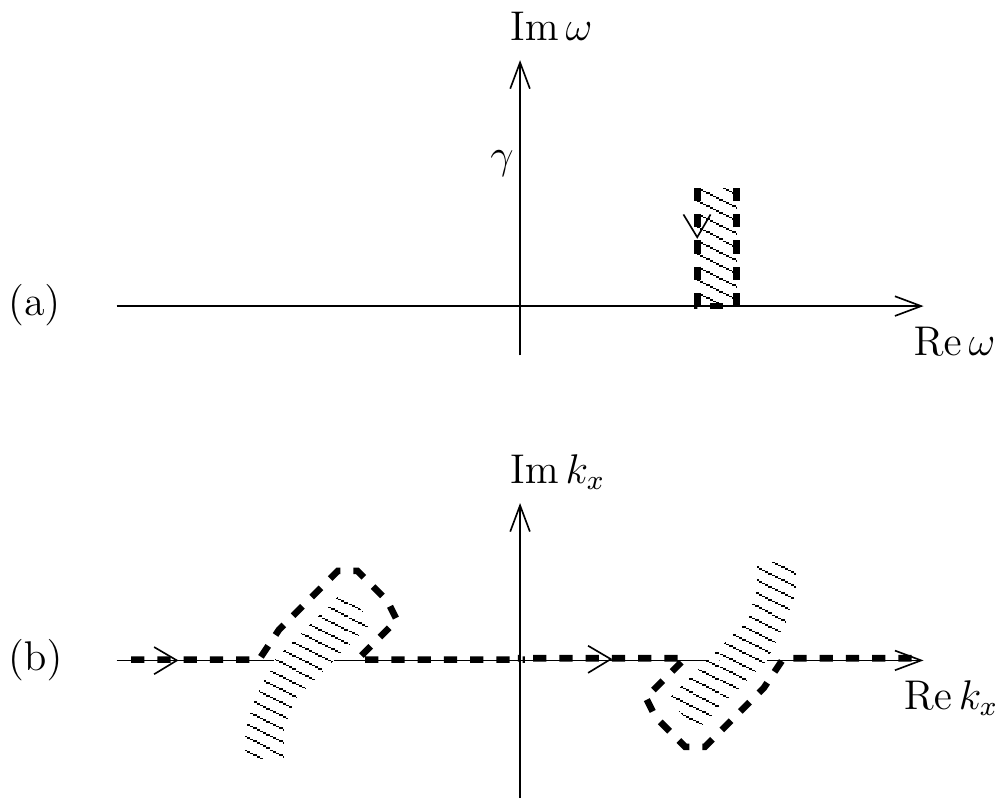}
\caption{Deformation in the $\omega$-domain. For each $k_x$ in the path in (b), the integration path in $D_\omega$ can be deformed (a).}
\label{fig:deformation2}
\end{figure}

Consider Fig. \ref{fig:deformation}a-b, showing the original integration paths in the $\omega$- and $k_x$-domains. For all $\omega$ in the indicated domain $D_\omega$, the branch points of $k_z$, i.e., $k_x=\pm\sqrt{\epsilon\mu}\omega/c$, are located in the domain $D_{k_x}$. Now, consider the short piece of the integration path that lies in $D_\omega$. For these $\omega$ values, the idea is to deform the corresponding $k_x$ integration path, as shown in Fig. \ref{fig:deformation}c. This can safely be done, since $k_z(k_x,\omega)$ and therefore $E(k_x,z,\omega)$ are analytic wrt. $k_x$ away from the branch cuts.

\begin{figure}
\centering \includegraphics[width=8cm]{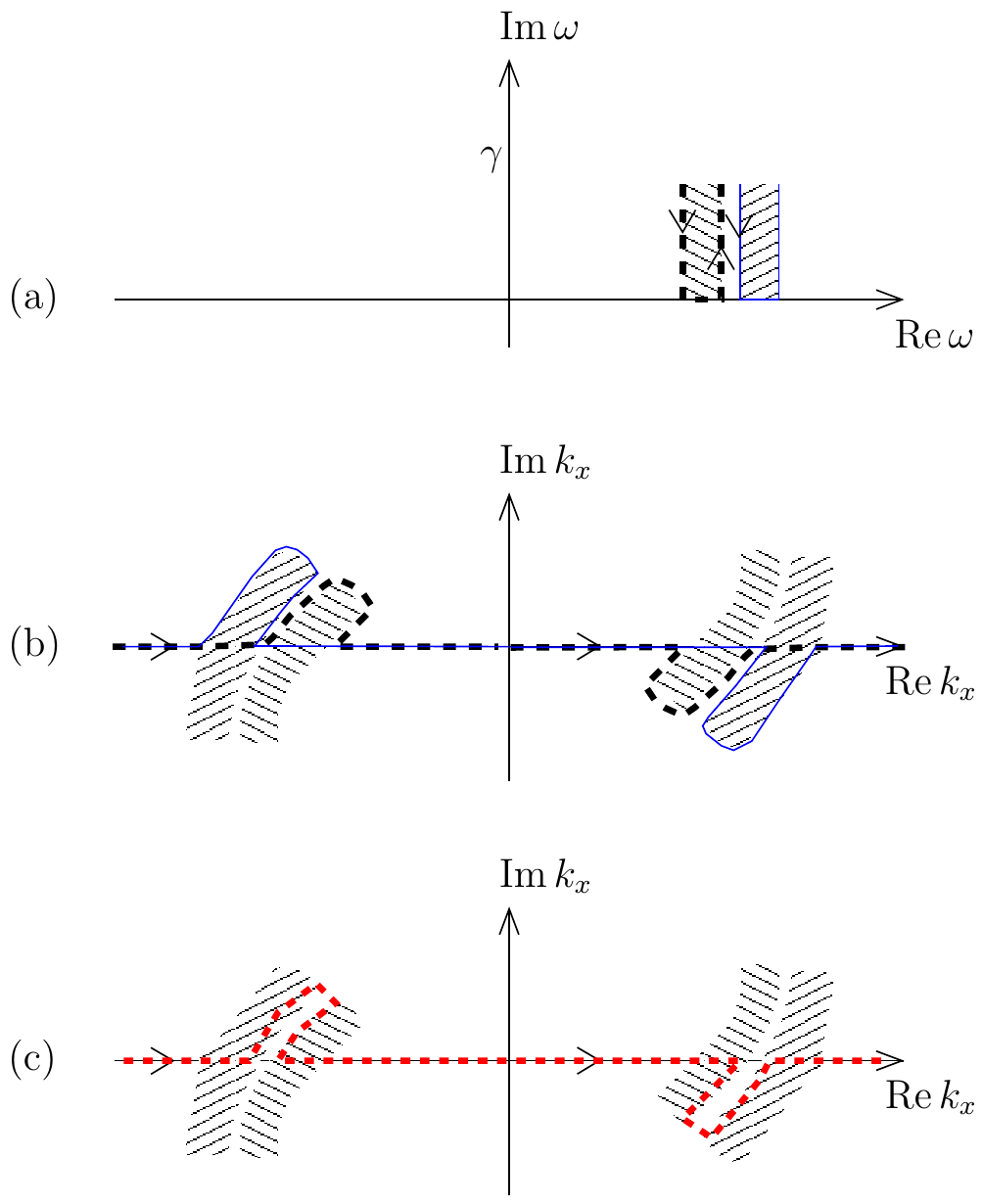}
\caption{Deformation of two neighboring pieces of the $\omega$-integration curve (dashed black and solid blue) (a) and the associated $k_x$-integration curves (b). For $\omega$-values along the vertical integration curves between the neighboring domains in (a), one can use a common $k_x$-integration curve (c).}
\label{fig:deformation3}
\end{figure}

The next step is to interchange the order of integration. For each $k_x$ in the path in Fig. \ref{fig:deformation2}b, we can deform the short piece of the $\omega$-path, obtaining the path in Fig. \ref{fig:deformation2}a. Repeating the procedure for two neighboring pieces of the $\omega$-integration curve, we obtain the situation in Fig. \ref{fig:deformation3}, generally with two different integration curves in the $k_x$-domain. In simple situations like the one in the figure, we could use a single, common integration curve in the $k_x$-domain, for both pieces in the $\omega$-domain. In general, to get rid of the vertical integration curves between the two domains in Fig. \ref{fig:deformation3}a, we must require the existence of a common integration curve in the $k_x$-domain detouring the interface between the neighboring domains (Fig. \ref{fig:deformation3}c). If this is always the case, we can continue the deformation in the $\omega$-domain until the integration curve coincides with the real axis:
\begin{equation}\label{invfour}
\mathcal E(x,z,t) = \frac{1}{2\pi}\int_{-\infty}^{\infty} E(x,z,\omega)\exp(-i\omega t)\diff\omega,
\end{equation}
where
\begin{equation}\label{freqdomainfield}
E(x,z,\omega)=\frac{1}{2\pi}\int_{\kappa(\omega)} E(k_x,z,\omega)\exp(ik_xx)\diff k_x.
\end{equation}
Here $\kappa(\omega)$ is the deformed path in the $k_x$-domain, for each $\omega$. Since $\im\omega=0$ in \eqref{invfour}, the resulting field will not diverge with time. Thus in these situations, there are no absolute instabilities, and \eqref{freqdomainfield} can be interpreted as the usual frequency-domain field for real $\omega$. The possible appearance of complex $k_x$'s in the integration path $\kappa(\omega)$, means that the field may grow with $x$.

We have required the existence of a common $k_x$-integration curve for any two neighboring $\omega$'s. To this end, consider the trajectories of $k_z$'s branch points, $k_x=\pm\sqrt{\epsilon\mu}\omega/c$, as we reduce $\im\omega$ from $\gamma$ to zero. It is necessary that for two neighboring values of $\re\omega$, these two trajectories will become arbitrarily close as the two $\re\omega$'s approach each other. A sufficient condition for this is that $\sqrt{\epsilon\mu}$ is analytic for $\im\omega\geq 0$. 

We have also required that $\epsilon\mu$ be zero-free for $\im\omega\geq 0$. While even order zeros give analytic square root, they induce another problem: At the zero the two branch points in the $k_x$-domain coincide so the integration curve get ``stuck''.

The frequency-domain field $E(x,z,\omega)$ is related to the physical, time-domain field in the so-called monochromatic limit. From \eqref{invfour},
\begin{equation}\label{monlimittimedomain}
\mathcal E(x,z,t)=\frac{1}{2\pi}\int_{-\infty}^\infty \frac{E(x,z,\omega)}{V(\omega)}V(\omega)\exp(-i\omega t)\diff\omega,
\end{equation}
where $E(x,z,\omega)/V(\omega)$ is the transfer function from the excitation $V(\omega)$ to the resulting field $E(x,z,\omega)$, as given by \eqref{Ezpos}. Note that $V(\omega)$ is a factor in $E(x,z,\omega)$, so the transfer function is independent of $V(\omega)$. We can for example take a unit-step modulated complex exponential as the excitation: 
\begin{equation}\label{unitstepsine}
v(t)=H(t)\exp(-i\omega_1t), \quad H(t)=\begin{cases} 0, & t<0 \\ 1, & t>0\end{cases},
\end{equation}
with Laplace transform
\begin{equation}\label{laplexcit}
V(\omega)=\frac{i}{\omega-\omega_1}.
\end{equation}
The inverse transform \eqref{monlimittimedomain} can be found with the residue theorem, by closing the contour by a large semicircle in the lower half-plane:
\begin{align}\label{monlimitres}
\mathcal E(x,z,t) &= \left[\frac{E(x,z,\omega)}{V(\omega)}\exp(-i\omega t)\right]_{\omega=\omega_1}\\
&+\text{transients}(t).\nonumber
\end{align}
Here the term transients$(t)$ is a result of the integration around all singularities and cuts in the lower half-plane, and will decay exponentially. For later use, we define the monochromatic limit $\lim_{\omega_1}\mathcal E(x,z,t)$ as the field when the excitation is given by \eqref{unitstepsine}, and for sufficiently large $t$ such that the transients can be ignored:
\begin{equation}\label{monlimitdef}
\lim_{\omega_1} \mathcal E(x,z,t) = \frac{E(x,z,\omega_1)}{V(\omega_1)}\exp(-i\omega_1 t),
\end{equation}
valid when $\epsilon\mu$ is analytic and zero-free for $\im\omega\geq 0$. Even though the monochromatic limit exists in principle, in some situations (media with large gain and large $x$ or $z$) the transients may be extremely strong, which means it may take very long time before they have died out.

\begin{figure}
\centering \includegraphics[width=8cm]{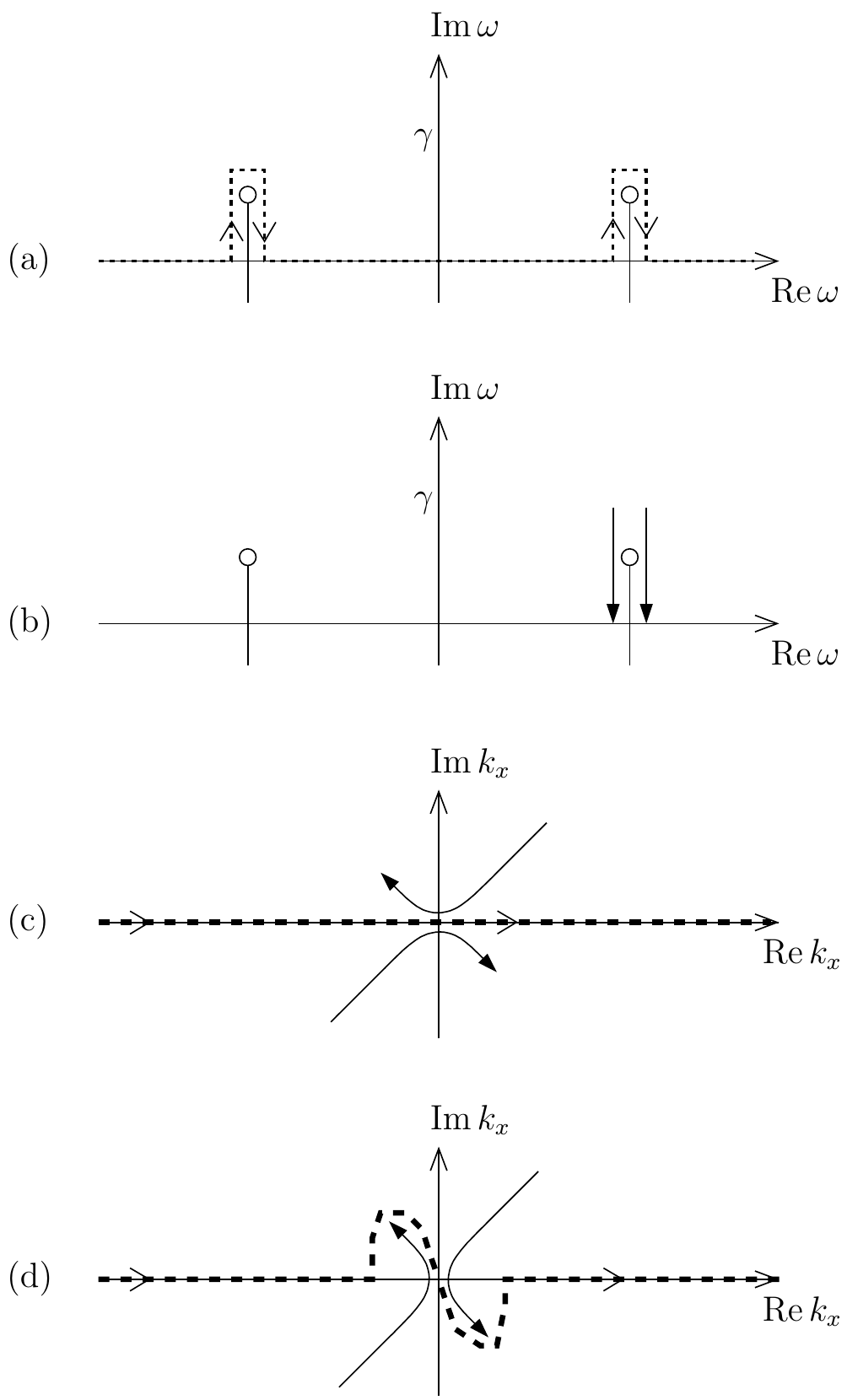}
\caption{Deformed integration paths (dashed) when $\sqrt{\epsilon\mu}$ has branch cuts in the upper half-plane. The branch points of $\sqrt{\epsilon\mu}$ are shown by open circles in (a); the cuts go vertically towards $-i\infty$. As $\im\omega$ is reduced from $\gamma$ to zero along the left and right arrows in (b), the corresponding trajectories of $k_x=\pm\sqrt{\epsilon\mu}\omega/c$ are shown by solid lines in (c) and (d), respectively.}
\label{fig:deformation4}
\end{figure}

We now consider the more complicated situation where $\epsilon\mu$ is not analytic or zero-free everywhere in the upper half-plane $\im\omega>0$. For concreteness we assume $\epsilon\mu$ has two simple zeros but is analytic otherwise. Then $\sqrt{\epsilon\mu}$ has branch cuts, which we take to be vertical towards $-i\infty$. Since $\sqrt{\epsilon\mu}$ is analytic everywhere in the upper half-plane except at the branch cuts, we can use the procedure above to deform the integration paths, leading to the $\omega$-integration curve depicted in Fig. \ref{fig:deformation4}a. It is natural to try to deform also the remaining detours, to reach the real $\omega$-axis everywhere. To this end we let $\im\omega$ be reduced from $\gamma$ to zero, on the left-hand side and right-hand side of $\sqrt{\epsilon\mu}$'s branch cut (Fig \ref{fig:deformation4}b). The corresponding trajectories of $k_x=\pm\sqrt{\epsilon\mu}\omega/c$ are shown in Figs. \ref{fig:deformation4}c and \ref{fig:deformation4}d, respectively. Apparently, the result of the integration in Fig. \ref{fig:deformation4}c differ from that of Fig. \ref{fig:deformation4}d, so the integrations up and down in Fig. \ref{fig:deformation4}a generally do not cancel. As a result the detours cannot be omitted. The necessary presence of complex frequencies $\exp(-i\omega t)$ with $\im\omega>0$ means that the field will diverge with time, even at a fixed point in space. This means that the field cannot be interpreted at real frequencies as in \eqref{monlimitres}; we have an absolute instability. 

\section{Plane wave limit}\label{sec:realfreqplanewave}
We have seen that when there are no absolute instabilities, it is possible to move the inverse Laplace transform path down to the real axis, enabling interpretation of the fields \eqref{EH} and \eqref{EHm} for real frequencies. However, considering active media, this has come at a price: The integration curve in $k_x$ must be deformed to include complex values of $k_x$. As will be demonstrated shortly, this means that it is not necessarily possible to approach the plane wave limit any longer.

Consider an excitation in the form $u(x)v(t)$, with transform $U(k_x)V(\omega)$. The function $v(t)$ could be given by \eqref{unitstepsine}, while $u(x)$ could be e.g. one of the following alternatives:
\begin{subequations}
\begin{align}
u_1(x)&=\text{beam}(x/\sigma)\exp(iK_xx), \\
u_2(x)&=\exp(-x^2/2\sigma^2) \exp(iK_xx).
\end{align}
\end{subequations}
Here beam$(x/\sigma)$ stands for a function which vanishes for $|x|>\sigma$, is smooth for $|x|<\sigma$, and beam$(0)=1$. Both alternatives represent a beam of thickness $\sim \sigma$ and a bundle of $k_x$'s around the central transversal wavenumber $K_x$. The wavenumber spectra of the excitations are given by
\begin{subequations}\label{Usources}
\begin{align}
U_1(k_x)&=\sigma\,\text{Beam}[\sigma(k_x-K_x)],\\
U_2(k_x)&=\sqrt{2\pi}\sigma\exp[-\sigma^2(k_x-K_x)^2/2].
\end{align}
\end{subequations}
Here Beam$(k_x)$ is the Fourier transform of beam$(x)$. Both spectra $U_1(k_x)$ and $U_2(k_x)$ are entire functions in $k_x$. For real $k_x$ and $K_x$ both functions can be thought of as $2\pi\cdot\delta(k_x-K_x)$ in the limit $\sigma\to\infty$. However, the two of them are fundamentally different in the sense that the first goes slowly to zero compared to the second. Indeed, if the $m$th derivative of beam$(x)$ is nonzero at the endpoints, while the lower order derivatives vanish, the asymptotic behavior of $U_1(k_x)$ for large $|k_x|$ is 
\be\label{asymptU1}
|U_1(k_x)|\sim \frac{\exp(|\im k_x|\sigma)}{|k_x|^{m+1}\sigma^m},
\ee
as can be proved using integration by parts. A similar result is valid for smooth functions with support $[-\sigma,\sigma]$ (so-called bump functions), except that the decay along the real $k_x$ axis is faster than 1/polynomial and slower than an exponential.

We will now consider $\lim_{\sigma\to\infty}\lim_{\omega_1}\mathcal E(x,z,t)$. This limit can be realized as follows. We pick an excitation width $\sigma$ and perform the experiment, waiting a sufficiently long time such that the electric field has reached the monochromatic limit. Next we pick a larger $\sigma$ and repeat the experiment, waiting a sufficiently long time (possibly longer than the first time) until the field has reached the monochromatic limit. After repeating the experiment several times, with increasing $\sigma$, the field will tend to $\lim_{\sigma\to\infty}\lim_{\omega_1}\mathcal E(x,z,t)$.   

The monochromatic limit is given by \eqref{monlimitdef}, so we need to consider the limit $\sigma\to\infty$ in \eqref{freqdomainfield}, expressed at the real excitation frequency $\omega_1$. To this end, we have assumed that the monochromatic limit exists (no absolute instabilities), i.e., $\epsilon\mu$ has no poles or zeros in the upper half-plane $\im\omega\geq 0$. The integration path $\kappa(\omega_1)$ in the $k_x$-plane, such as that in Fig. \ref{fig:deformation2}b, involves complex $k_x$. Surprisingly, now the limit $\sigma\to\infty$ does not necessarily exist, as $U_1(k_x)$ diverges for complex $k_x$. In fact, this will always be the case in practice, since the excitation necessarily must have finite support to be realizable.

However, from a theoretical perspective it is quite common to consider gaussian beams or excitations, so it is interesting to consider the possibility $U_2(k_x)$. Surprisingly, even though the plane wave limit $\sigma\to\infty$ did not exist when using $U_1(k_x)$, it may exist when using $U_2(k_x)$, since the gaussian tends to zero provided $|\im k_x| < |\re k_x-K_x|$. Thus, when the detours of the $k_x$-integration curve are not too far away from the real axis, or too close to the excitation wavenumber $K_x$, we can take the plane wave limit using a gaussian excitation, but not a finite-support excitation. When the limit exists, we can write
\begin{equation}\label{planwavetilde}
E(x,z,\omega_1) =\frac{\tilde E(K_x,z,\omega_1)}{U(K_x)}\exp(iK_xx),
\end{equation}
for some function $\tilde E(K_x,z,\omega_1)$, expressing the field with a single wavenumber $K_x$. For most media the part of the integration in \eqref{freqdomainfield} along the real axis is one way, which means that $\tilde E(K_x,z,\omega_1)=E(K_x,z,\omega_1)$. For certain, very special media, as we will see in Sec.  \ref{simrefrindsec}, the integration along part of the real axis will give rise to one more term in $\tilde E(K_x,z,\omega_1)$. Eq. \eqref{planwavetilde} means that the physical time domain field in the monochromatic limit will tend to
\begin{equation}\label{planewavemonochromatic}
\lim_{\sigma\to\infty}\lim_{\omega_1} \mathcal E(x,z,t) = \frac{\tilde E(K_x,z,\omega_1)}{U(K_x)V(\omega_1)}\exp(iK_xx-i\omega_1 t),
\end{equation}
as the width of the gaussian tends to infinity.

The peculiar divergence discussed above, can be interpreted as follows. For certain frequencies $\omega$ and wavenumbers $\pm k_x$, the longitudinal wavenumber $k_z$ becomes zero. These modes correspond to side waves, which propagate in the $\pm x$-direction. If the medium is gainy, and the excitation extends over all $x$'s, the field at an observation point $x$ may diverge since the side waves propagate an unlimited distance before reaching the point. For the finite-support excitation $u_1(x)$, as $\sigma$ increases, side waves will have the chance to propagate a larger distance before reaching the observation point; thus we expect an exponential growth. At the same time, the excitation $U_1(k_x)$ at the particular $k_x$ associated with the side wave becomes weaker, but only as $\propto\sigma^{-m}$. For the gaussian excitation $u_2(x)$, an increased $\sigma$ will again give rise to an exponential growth as a result of the increased distance; however, the excitation itself at the particular $k_x$ associated with the side wave, may be much weaker due to the factor $\exp[-\sigma^2(k_x-K_x)^2/2]$.

We now consider $\lim_{\omega_1}\lim_{\sigma\to\infty}\mathcal E(x,z,t)$. This order of limits is more difficult to realize that the opposite order, but can be approached by measuring the time response $\mathcal E(x,z,t)$ for a fixed time interval, repeating the experiment for increasing $\sigma$. After convergence, the time interval is shifted to later times, and the series of experiments is repeated, etc.

Mathematically, $\lim_{\omega_1}\lim_{\sigma\to\infty}\mathcal E(x,z,t)$ is found most easily by taking the limit $\sigma\to\infty$ in \eqref{invlaplfour}.
Since only real $k_x$'s are involved in the integral, the limit $\sigma\to\infty$ always exists, which leads to
\begin{equation}\label{invlaplrev}
\lim_{\sigma\to\infty}\mathcal E(x,z,t) = \frac{1}{2\pi}\int_{i\gamma-\infty}^{i\gamma+\infty} \frac{E(K_x,z,\omega)}{U(K_x)}\exp(iK_xx-i\omega t)\diff\omega.
\end{equation}
Equation \eqref{invlaplrev} has the disadvantage that it is expressed using complex frequencies. We would like to be able to set $\gamma=0$ in \eqref{invlaplrev} for interpretation at real frequencies. If the integrand is analytic for $\im\omega\geq 0$, we can move the integration path to the real axis. However, as we will see below, this is not always the case, not even for media with analytic and zero-free $\epsilon\mu$ for $\im\omega\geq 0$. Since $K_x$ is fixed, we must require that $\sqrt{\epsilon\mu\omega^2/c^2-K_x^2}$ is analytic in the upper half-plane $\im\omega\geq 0$, to avoid absolute instabilities. Although this can happen, it is not very common; for $K_x\neq 0$ it is not even the case for conventional, weak gain media \cite{skaar06b,grepstad11}: For such media, there is a branch point slightly above the real $\omega$-axis, corresponding to a side wave with $K_z=0$. For plane wave excitations, this side wave propagates an infinite distance along the $x$-axis, thus picking up an infinite amount of gain.

This type of absolute instability is somewhat artificial, since it is induced by an excitation of infinite width. For the case with finite $\sigma$ we have seen that the instability is only convective, as long as the medium has analytic and zero-free $\epsilon\mu$ for $\im\omega\geq 0$. This makes sense intuitively, since for finite $\sigma$, the side wave has only propagated a finite distance from the excitation to a fixed observation point.

In other words: If $\epsilon\mu$ is analytic and zero-free for $\im\omega\geq 0$, but $\sqrt{\epsilon\mu\omega^2/c^2-K_x^2}$ is not analytic there (which is the case e.g. for a weak inverted Lorentzian and $K_x\neq 0$),
\begin{equation}
\lim_{\omega_1} \mathcal E(x,z,t) = \text{finite} 
\end{equation}
for any finite $\sigma$, while
\begin{equation}
\lim_{\omega_1}\lim_{\sigma\to\infty} \mathcal E(x,z,t) = \infty. 
\end{equation}
However,
\begin{equation}
\lim_{\sigma\to\infty}\lim_{\omega_1} \mathcal E(x,z,t),
\end{equation}
on the other hand, is dependent on the manner in which the plane wave limit is taken. If it is taken using an excitation $U_1(k_x)$ of finite support, it is infinite, but if it is taken using a gaussian $U_2(k_x)$, it is finite provided $|\im k_x| < |\re k_x-K_x|$ along the integration detour. The gaussian excitation $u_2(x)$ is somewhat unphysical, as it requires an infinitely wide source even for finite $\sigma$. Even though the gaussian excitation is unphysical, the fact that it makes it possible to take the plane wave limit is interesting. It tells us that the growing side waves in a gain medium may be reduced by making the source sufficiently smooth, and will disappear in the limit of a perfect gaussian.

Remarkably, and less intuitively, for certain media with absolute instabilities for finite $\sigma$ (meaning that $\epsilon\mu$ is not analytic and zero-free everywhere in the upper half-plane), it is possible to eliminate the absolute instabilities by letting $\sigma\to\infty$. Indeed, if $\sqrt{\epsilon\mu\omega^2/c^2-K_x^2}$ is analytic for $\im\omega>0$ while $\epsilon\mu$ is not analytic and zero-free,
\begin{equation}\label{st1}
\lim_{\omega_1} \mathcal E(x,z,t) = \infty, 
\end{equation}
for any $\sigma$, while
\begin{equation}\label{st2}
\lim_{\omega_1}\lim_{\sigma\to\infty} \mathcal E(x,z,t) = \text{finite}. 
\end{equation}
For example, this happens for media for which $\epsilon\mu\omega^2/c^2-K_x^2$ has no zeros in the upper half-plane $\im\omega>0$, while $\epsilon\mu$ has two simple zeros there. Such a medium is suggested in Ref. \cite{grepstad11}. Equations \eqref{st1} and \eqref{st2} can be interpreted as follows. Consider the field $\mathcal E(x,z,t)$ when $\sigma$ and $t$ are finite. As $\sigma$ is made larger, the unstable mode with $k_z=0$ is excited more weakly. Thus a larger $t$ can be tolerated before $\mathcal E(x,z,t)$ gets large. If  $\sigma\to\infty$ first, we can let $t$ be infinite as well, without getting an infinite field. Thus the monochromatic limit exists.

We conclude this section by noting that the monochromatic and plane wave limits are far from trivial in gain media. Although it can be argued that these limits are unphysical, since infinite experiment durations or infinite beam thicknesses cannot exist, they provide valuable intuition for experiments with wide beam excitations, or long duration. Apparently, different results may be obtained dependent on the wideness of the excitation and the duration of the experiment.

\section{Media}\label{sec:examples}
The general method from the previous sections is now applied to analyze a wide range of media of interest, starting with simple passive and active media, and ending with novel classes of active media.

\subsection{Passive media}\label{sec:passive}
Passive media are simple to analyze, due to the absence of instabilities. Fourier analysis is therefore sufficient, and the Fourier components wrt. $k_x$ and $\omega$ can be interpreted straightforwardly. Although these facts are well known, it is useful to demonstrate the formalism before moving on to more complex cases.

\begin{figure}
\centering \includegraphics[width=8cm]{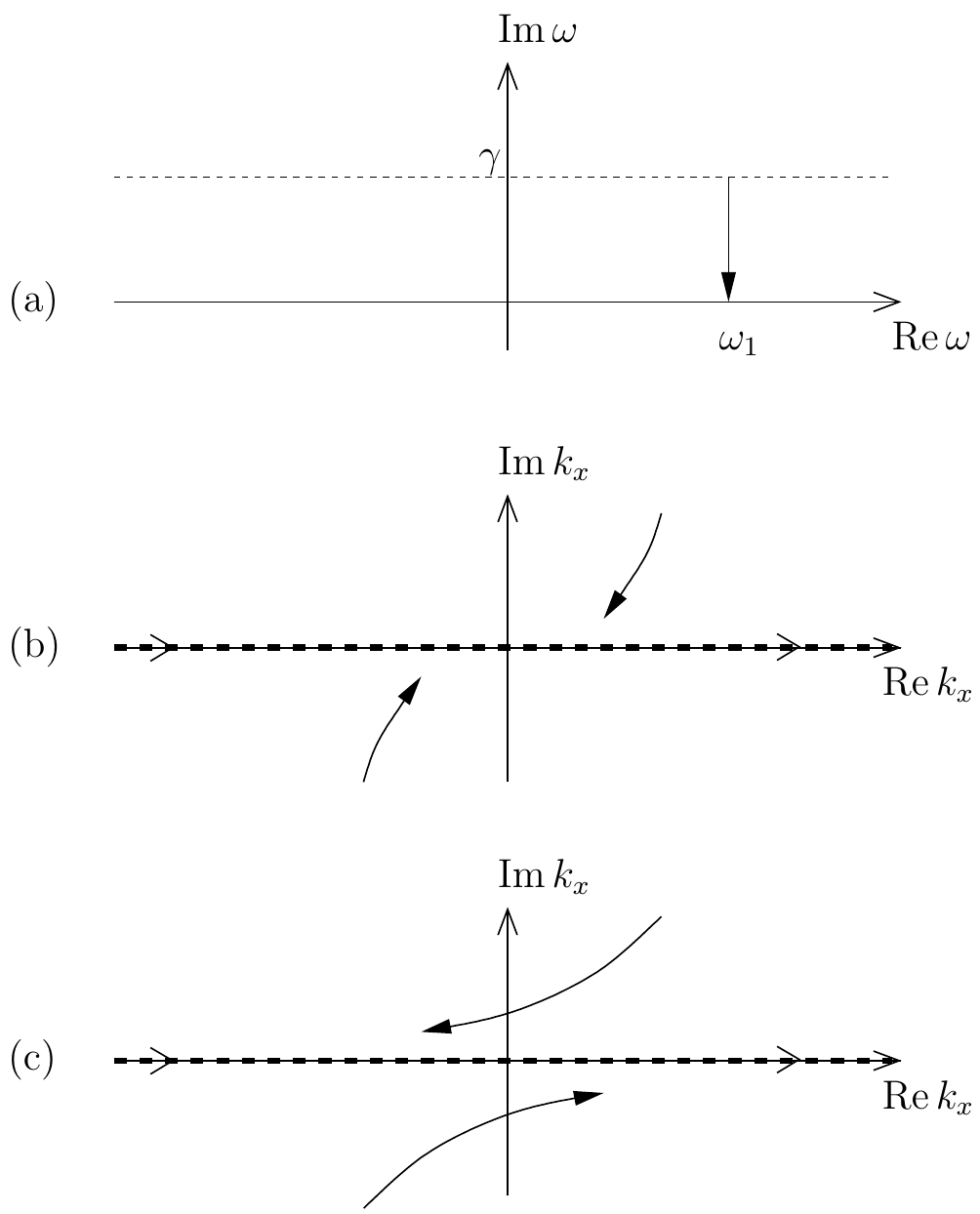}
\caption{As $\im\omega$ is reduced from $\gamma$ to zero (a), $k_z$'s branch points, $k_x=\pm\sqrt{\epsilon\mu}\omega/c$, moves along the trajectories in (b) for a passive, positive refractive medium, and (c) for a passive, negative refractive medium.}
\label{fig:passive}
\end{figure}

A passive medium has $\im\epsilon(\omega)>0$, $\im\mu(\omega)>0$, and $\im n(\omega)>0$ for $\omega>0$. Here $n(\omega)=\sqrt{\epsilon\mu}$ is the refractive index, which is analytic in the upper half-plane \cite{nussenzveig}. Due to odd symmetry of these functions, $\im n\omega/c\geq 0$ for all real $\omega$. Since $\im n\omega/c$ is a harmonic function \cite{ahlfors}, it takes its minimum on the real axis; thus $\im n\omega/c\geq 0$ in the closed upper half-plane. It follows that $k_z$'s branch points, $k_x=\pm n\omega/c$, do not cross the real $k_x$-axis as we reduce $\im\omega$ towards zero. In Fig. \ref{fig:passive} we show two different possibilities; a passive medium which will turn out to show positive refraction (b), and a passive medium with negative refraction (c). Clearly, in both cases we can integrate along the real $\omega$ and $k_x$ axes, and the monochromatic and plane wave limits may be taken, leading to fields with frequency $\omega_1$ and wavenumber $K_x$. The resulting $K_z$ shows the behavior of the wave in the medium. 

We can find the sign of $K_z$ by tracing $\arg k_z$ as $k_x$ decreases from $+\infty$ to $K_x$. For $k_x\to+\infty$, $k_z\to +i k_x$ (see Appendix \ref{sec:fubini}). As $k_x$ decreases, consider $k_z^2=\epsilon(\omega_1)\mu(\omega_1)\omega_1^2/c^2-k_x^2$, with the two zeros shown by the solid arrow ends in Figs. \ref{fig:passive}b-c. Now, $k_z^2$ picks up phase from the two zeros, but very little if $k_x$ is in the regime far away from the zeros. Since $k_z(k_x,\omega)$ is continuous in $k_x$ away from the branch cuts, it follows that $K_z=k_z(K_x,\omega_1)\approx iK_x$ in the regime far to the right of the zeros, corresponding to an evanescent behavior in the total internal reflection regime of large $K_x$. So far, we have not invoked the properties of the medium; in other words, the result is valid for all media and situations where the monochromatic and plane wave limits exist.

As $K_x$ becomes smaller, we must consider the two passive media separately. For the positive refractive medium (Fig. \ref{fig:passive}b), since the right-hand zero is above the real $k_x$-axis, as we pass it on the way from large $k_x$ to small $k_x$, the phase $\arg k_z^2$ reduces from $\pi$ through $\pi/2$ towards $\arg\{\epsilon(\omega_1)\mu(\omega_1)\}$. Again, since $k_z(k_x,\omega)$ is continuous in $k_x$ away from the branch cuts, it follows that $\arg k_z(k_x,\omega_1)$ goes from $\pi/2$ through $\pi/4$ towards the small number $\arg\{\epsilon(\omega_1)\mu(\omega_1)\}/2$. Thus, as expected, we obtain a damped, propagating wave with wave vector directed away from the source.

For the negative refractive medium (Fig. \ref{fig:passive}c), the right-hand zero is below the real $k_x$-axis. Thus we find that $\arg k_z^2$ increases from $\pi$ to almost $2\pi$, and therefore, $\arg k_z$ increases from $\pi/2$ to almost $\pi$. In other words, $K_z$ will be close to a negative number (negative refraction) in the regime of small $K_x$.

\subsection{Weak gain medium}\label{sec:weakgain}
\begin{figure}
\centering \includegraphics[width=8cm]{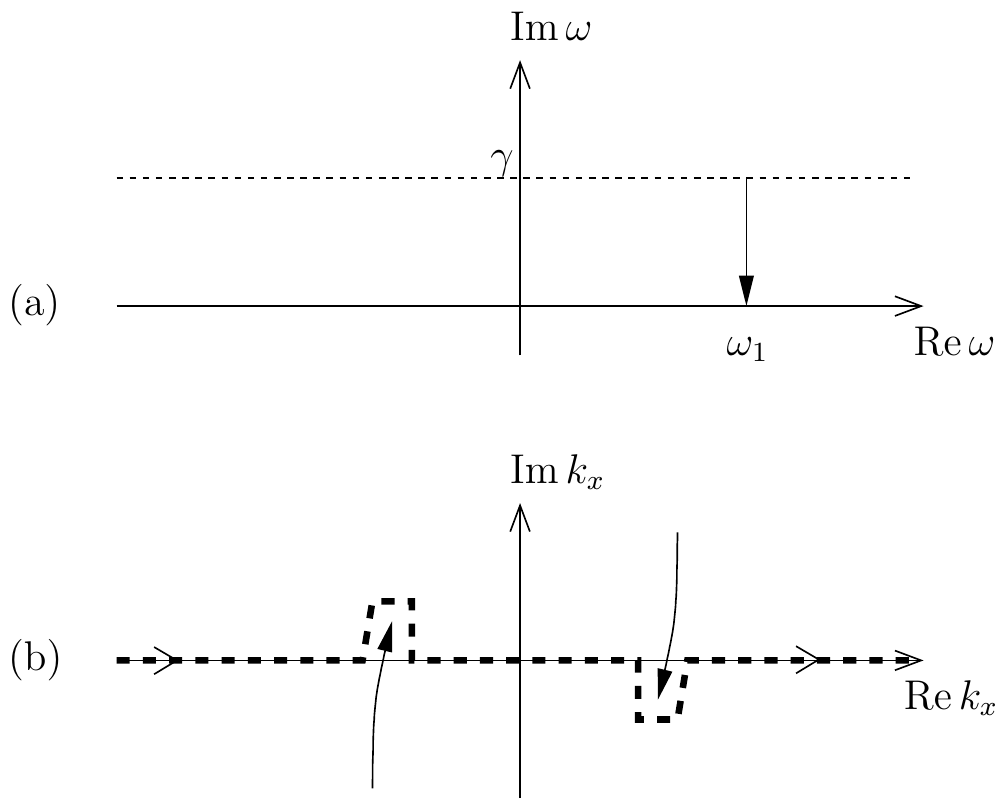}
\caption{As $\im\omega$ is reduced from $\gamma$ to zero (a), $k_z$'s branch points, $k_x=\pm\sqrt{\epsilon\mu}\omega/c$, moves along the trajectories in (b) for a weak gain medium. The integration path in the $k_x$-domain must detour around these branch points.}
\label{fig:weakgain}
\end{figure}

We now consider a weak gain medium, or conventional gain medium, with $|\im\epsilon| \ll 1$ and $|\im\mu| \ll 1$ for all frequencies, and weak dispersion. For example, we can consider a nonmagnetic medium with $\epsilon(\omega)=1+\chi(\omega)$, where $\im\chi(\omega)$ is negative at the observation frequency, and $|\chi(\omega)|\ll 1$ for all $\omega$. When we reduce $\im\omega$ as in Fig. \ref{fig:weakgain}a, the branch points $k_x=\pm\sqrt{\epsilon\mu}\omega/c$ move according to Fig. \ref{fig:weakgain}b. Thus, to be able to express the integral \eqref{invlaplfourGamma} with real frequencies $\omega$, it is necessary to deform the $k_x$-integration with detours. These detours are result of the fact that the system supports amplifying side waves with $k_x=\pm\sqrt{\epsilon\mu}\omega/c$.

Having taken the monochromatic limit, we consider the possibility of approaching plane waves. According to the discussion in Sec. \ref{sec:realfreqplanewave}, the limit $\sigma\to\infty$ does not exist when using excitation profiles of finite support; then the side waves will diverge. However, for the gaussian excitation profile $u_2(x)$, and provided $|\im \sqrt{\epsilon\mu}\omega_1/c| < |\re \sqrt{\epsilon\mu}\omega_1/c-K_x|$, we can take the plane wave limit, since then the side waves are very weakly excited. By tracing $\arg k_z$ as $k_x$ is reduced from $\infty$ (as in Sec. \ref{sec:passive}), we still obtain $K_z\approx iK_x$ in the total internal refraction regime of large $K_x$. Thus the behavior remains approximately evanescent there, in agreement with earlier predictions \cite{kolokolov99} and finite difference time domain (FDTD) simulations \cite{willis08}. FDTD simulations solve Maxwell's equations directly in the time domain, and thus provide an independent verification of the theory. For small $K_x$, since we have passed the zero from below, we get $\arg K_z \approx \arg\{\epsilon(\omega_1)\mu(\omega_1)\}/2$. This represents a weakly amplified wave, a result that is well documented with numerous experiments and simulations.

As an alternative, we can take the plane wave limit while keeping the Bromwich integration path at $\im\omega=\gamma$, leading to a single wavenumber $K_x$. Then we can deform the Bromwich path towards the real axis; however, there will be branch points close to $\omega=K_xc$, above the real axis. This means that the system supports absolute instabilities, and that the real frequencies are not meaningful in general. The absolute instabilities are again a result of diverging side waves, being excited infinitely far away from the observation point. However, as shown in Ref. \cite{grepstad11}, as long as the excitation frequency $\omega_1$ is far away from $K_xc$, we can interpret the field as ``quasi-monochromatic'' up to a certain time, where the diverging side waves start to dominate.

\subsection{Non-magnetic negative index medium}
If the permittivity and permeability from the negative index medium in Sec. \ref{sec:passive} are denoted $\epsilon_\text{p}$ and $\mu_\text{p}$, we let the permittivity of an active, nonmagnetic medium be $\epsilon=\epsilon_\text{p}\mu_\text{p}$. Clearly, the behavior of the branch points and the integration paths becomes identical to that in Fig. \ref{fig:passive}c, and we get a negative refractive index at the frequency shown in the figure. This type of media was suggested in Ref. \cite{chen05} and analyzed in Ref. \cite{skaar06}. When a plane wave is normally incident from vacuum, a backward wave is excited in the medium, drawing energy from the medium and propagating energy towards the interface \cite{skaar06}. However, note that both the phase velocity and the Poynting vector point backwards, so the medium is fundamentally different from left-handed negative index media.  

The fact that this type of media exhibits negative refraction, has also been independently verified through time domain simulations, e.g. in Ref. \cite{geddes07}.

\subsection{Anti-evanescent medium}
Having analyzed previously known media with the $\omega$- and $k_x$-integration formalism, we now consider how the formalism can be used to predict novel classes of media. As we reduce $\im\omega$ from $\gamma$ to zero, the trajectories of $k_z$'s  branch points may be more complicated than in the previous examples.

\begin{figure}
\centering \includegraphics[width=9cm]{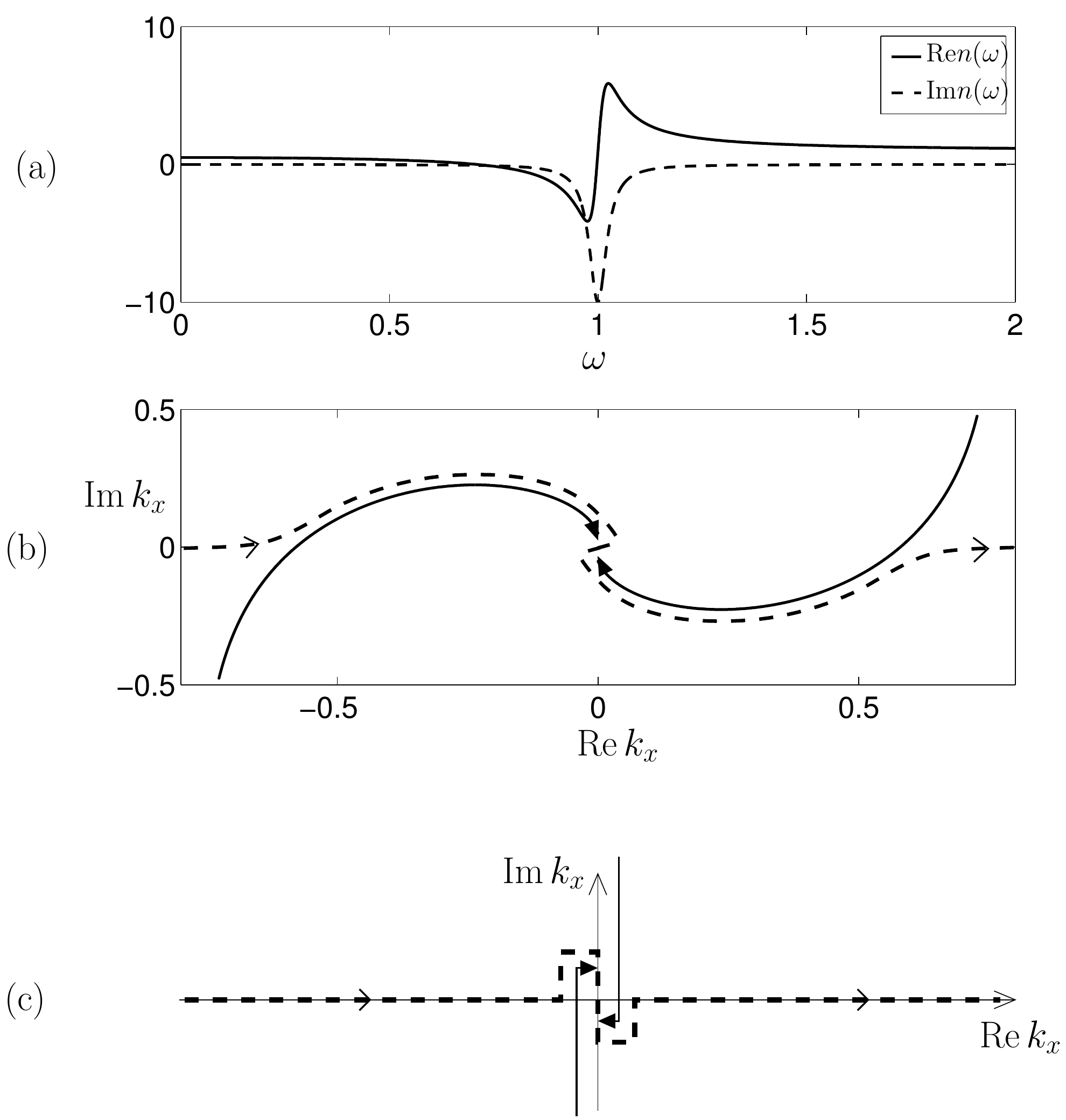}
\caption{Plot (a) shows $n(\omega)$ as given by \eqref{refrantiev}. Plot (b) shows the trajectories of $k_z$'s branch points, $k_x=\pm n(\omega)\omega/c$, as $\im\omega$ is reduced from $\gamma$ to zero, and $\re\omega=\omega_1$. The values for $\omega=\omega_1$ are shown with solid arrows. The branch cuts in the $k_x$-domain, for $\omega=\omega_1$, can be taken along the trajectories (b, solid lines); however, it is convenient to use analytic continuation to deform them into the solid lines shown in (c). The integration path in the $k_x$-domain (dashed) must detour around the branch cuts.}
\label{fig:antiev}
\end{figure}

Consider a medium with refractive index
\begin{equation}\label{refrantiev}
n(\omega)=1-\frac{F\omega_0^2}{\omega_0^2-\omega^2-i\Gamma\omega}, 
\end{equation}
and $F>0$ (see Fig. \ref{fig:antiev}a). This refractive index can be obtained e.g. by letting $\epsilon(\omega)=\mu(\omega)=n(\omega)$. Such Lorentzian resonances can be realized in metamaterials; however there are challenges associated with high gain (see Subsection \ref{simrefrindsec}). The same refractive index can be obtained by setting $\epsilon(\omega)=(n(\omega))^2$ and $\mu=1$. In the following we will for simplicity consider this nonmagnetic realization. 

Provided $F<1$, the zeros of $\epsilon(\omega)$ are located in the lower half-plane, so the medium does not support absolute instabilities. Hence we may consider the monochromatic limit. We take $F=0.5$ and $\Gamma=0.05\omega_0$, and consider the observation frequency $\omega_1=0.71\omega_0$, for which $\re n(\omega_1)=0$ and $\im n(\omega_1)=-i0.072$.

The trajectories of $k_z$'s branch points, $k_x=\pm n(\omega)\omega/c$, as $\im\omega$ is reduced from $\gamma$ to 0 while $\re\omega=\omega_1$, are shown in Fig. \ref{fig:antiev}b. For $\omega=\omega_1$ we can take the branch cuts along the solid lines in Fig. \ref{fig:antiev}c, and the integration path along the dashed line. We let the two branch cuts approach each other. Considering an incident wave from vacuum, we find with the help of \eqref{Ezposb} and \eqref{freqdomainfield}:
\begin{align}\label{antievfield}
& 2\pi\frac{E(x,z,\omega_1)}{V(\omega_1)} \\
& = \int_{-\infty}^{\infty}U(k_x)\frac{2k_{1z}e^{ik_{2z}z}}{k_{1z}+k_{2z}}e^{ik_xx}\diff k_x \nonumber\\
& + \int_{-k_\text{b}}^{k_\text{b}}U(k_x)\left(\frac{2k_{1z}e^{ik_{2z}z}}{k_{1z}+k_{2z}}-\frac{2k_{1z}e^{-ik_{2z}z}}{k_{1z}-k_{2z}}\right)e^{ik_xx}\diff k_x. \nonumber
\end{align}
Here the integration $\int_{-k_\text{b}}^{k_\text{b}}$ is along a vertical path from the lower to the upper branch point (indicated with solid arrows in Fig. \ref{fig:antiev}c); immediately to the right of the branch cuts.

To interpret \eqref{antievfield}, we note that $k_{2z}^2=n^2(\omega_1)\omega_1^2/c^2-k_x^2$ is negative for real $k_x$ and also along the vertical integration paths in Fig. \ref{fig:antiev}c. Since $k_{2z}\to +i k_x$ for $k_x\to+\infty$, $k_{2z}$ must be positive imaginary for real $k_x$ away from the branch cuts. Along the imaginary axis, however, $k_{2z}$ becomes negative imaginary, due to the presence of the right-hand branch cut. We choose an excitation $U(k_x)=U_1(k_x)$, with $K_x=0$ (normal incidence). Clearly, the plane wave limit does not exist, as the second integral in \eqref{antievfield} involves complex $k_x$'s for which $U_1(k_x)$ diverges as $\sigma\to\infty$. For a finite, though large $\sigma$, the field is dominated by the second integral in \eqref{antievfield}. As a result of the two terms of the second integral, the field contains a superposition of modes with both signs of $k_z$; evanescent ($\im k_z > 0$) and anti-evanescent ($\im k_z < 0$).

The situation is different if we take the plane wave limit before the monochromatic limit. If we still assume $K_x=0$, we have $K_{2z}=+n(\omega)\omega/c$. Both limits exist, and we end up with the monochromatic field amplitude
\begin{equation}\label{antievfield2}
\mathcal E(x,z,t) = \frac{2K_{1z}}{K_{1z}+K_{2z}}e^{iK_xx+iK_{2z}z-i\omega_1t}.
\end{equation}
For the medium in this example, $n(\omega)$ is negative imaginary at the observation frequency $\omega=\omega_1$. Thus we have an anti-evanescent behavior.

In other words, let the beam width $\sigma$ be fixed and finite. Then, after sufficiently long time, the field will be a superposition of evanescent and anti-evanescent modes. On the other hand, for $\sigma\to\infty$, and after a long time the field will be purely anti-evanescent.

\subsection{Simultaneous refractive index medium}\label{simrefrindsec}
In the previous example, we observed that the evanescent and anti-evanescent modes were excited simultaneously. We will now demonstrate a remarkable result; that there exist isotropic media exhibiting positive and negative refraction simultaneously.

\begin{figure}
\centering \includegraphics[width=8cm]{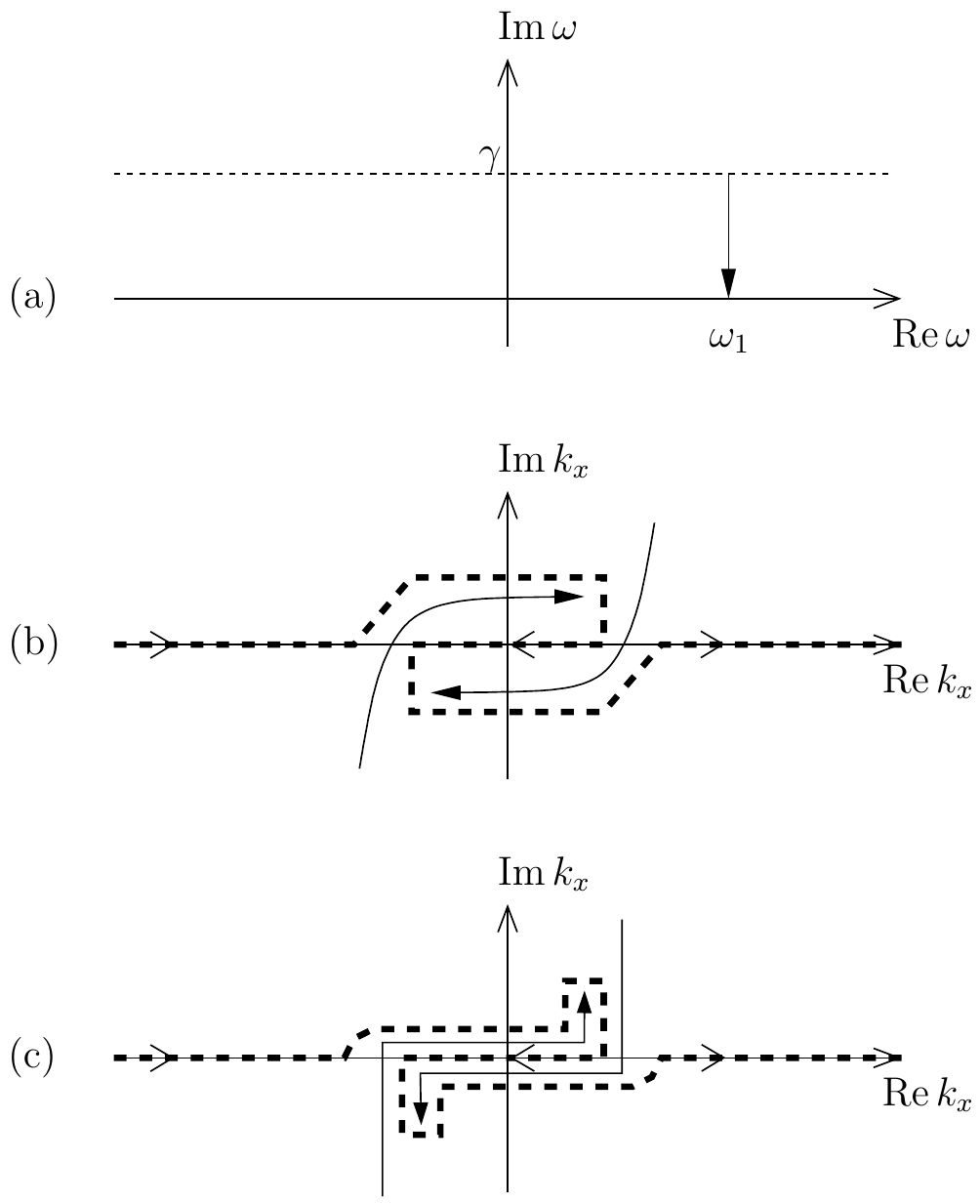}
\caption{As $\im\omega$ is reduced from $\gamma$ to zero (a), $k_z$'s branch points, $k_x=\pm\sqrt{\epsilon\mu}\omega/c$, move along the trajectories (b). By deforming branch cuts and integration paths, we get situation (c). In (b) the branch cuts are taken to be along the trajectories, while in (c) they are deformed into the solid lines.}
\label{fig:simrefr}
\end{figure}

Consider the example in Fig. \ref{fig:simrefr}. As $\omega$ approaches the real axis, the branch point in the first quadrant moves via the forth to the third quadrant. The integration path therefore becomes zigzag. We consider an incident wave from a passive medium (medium 1) to the medium under investigation (medium 2), and calculate the transmitted field using \eqref{Ezposb}. Using the integration path in Fig. \ref{fig:simrefr}c, this leads to
\begin{align}\label{simrefrfield}
& 2\pi\frac{E(x,z,\omega_1)}{V(\omega_1)} \\
& = \left(\int_{-\infty}^{-k_\text{b}}+\int_{k_\text{b}}^{\infty}\right)U(k_x)\frac{2\mu_2k_{1z}e^{ik_{2z}z}}{\mu_2k_{1z}+\mu_1k_{2z}}e^{ik_xx}\diff k_x \nonumber\\
& + \int_{-k_\text{b}}^{k_\text{b}}U(k_x)\left(\frac{4\mu_2k_{1z}e^{ik_{2z}z}}{\mu_2k_{1z}+\mu_1k_{2z}}-\frac{2\mu_2k_{1z}e^{-ik_{2z}z}}{\mu_2k_{1z}-\mu_1k_{2z}}\right)e^{ik_xx}\diff k_x \nonumber \\
& + \int_{\text{vertical detours}}U(k_x)\frac{2\mu_2k_{1z}e^{ik_{2z}z}}{\mu_2k_{1z}+\mu_1k_{2z}}e^{ik_xx}\diff k_x. \nonumber
\end{align}
In \eqref{simrefrfield} $k_\text{b}$ is the real part of the branch point in the first quadrant, and the last integral represents  all vertical integration paths in Fig. \ref{fig:simrefr}c, letting the up-and-down paths around a branch cut be infinitely close to each other. In the third line of \eqref{simrefrfield}, $k_{2z}$ is the value along the upper integration path, above both branch cuts.

Considering the observation frequency $\omega_1$ (monochromatic limit), we now take the plane wave limit $\sigma\to\infty$. Using the gaussian excitation $U_2(k_x)$, the limit exists provided $|\im k_x| < |\re k_x-K_x|$ on the integration path. Assuming $-k_\text{b}<K_x<k_\text{b}$, we end up with
\begin{align}\label{simrefrfieldplanew}
& 
\mathcal E(x,z,t) \\
& = \left(\frac{4\mu_2K_{1z}e^{iK_{2z}z}}{\mu_2K_{1z}+\mu_1K_{2z}}-\frac{2\mu_2K_{1z}e^{-iK_{2z}z}}{\mu_2K_{1z}-\mu_1K_{2z}}\right)e^{iK_xx-i\omega_1 t}. \nonumber 
\end{align}
With an excitation $u_1(x)$ of finite support, the limit would not exist; however, for bounded $x$ and $z$ we may come as close as we wish to the field \eqref{simrefrfieldplanew} by ensuring that the medium have branch points sufficiently close to the real $k_x$-axis. 

On the other hand, by taking the limit $\sigma\to\infty$ without taking the monochromatic limit, we get
\begin{align}\label{simrefr2}
& \mathcal E(x,z,t) = \frac{1}{2\pi}\int_{i\gamma-\infty}^{i\gamma+\infty} V(\omega) \nonumber\\ 
& \cdot\frac{2\mu_2K_{1z}}{\mu_2K_{1z}+\mu_1K_{2z}}\exp(iK_xx+iK_{2z}z-i\omega t)\diff\omega.
\end{align}
However, moving the integration path down to the real $\omega$-axis requires $K_{2z}$ to be analytic for $\im\omega\geq 0$. Even for weak gain media this will not be the case \cite{grepstad11}, except for the special case $K_x=0$. 

If $K_x=0$, and both $\sqrt{\epsilon\mu}$ and the Fresnel transmission coefficient are analytic for $\im\omega>0$, the integration path can in fact be moved down to the real $\omega$-axis. In the monochromatic limit we then get
\begin{align}\label{simrefrfieldplanew2}
& \mathcal E(x,z,t) = \frac{2\mu_2K_{1z}}{\mu_2K_{1z}+\mu_1K_{2z}}e^{iK_xx+iK_{2z}z-i\omega_1 t}, 
\end{align}
with $K_{2z}=+n(\omega_1)\omega_1/c$. This differs from \eqref{simrefrfieldplanew}, and once again the two orders of the monochromatic and plane wave limits yield different results.   

In other words, consider the case $K_x=0$, for a sufficiently large, but finite $\sigma$. In the monochromatic limit $t\to\infty$, the field will then be approximately given by \eqref{simrefrfieldplanew}, i.e. a superposition of waves with wavenumber $+K_{2z}$ and $-K_{2z}$ in the $z$-direction. However, if $\sigma\to\infty$ first, the monochromatic limit leads to a plane wave propagating in the $z$-direction, with wavenumber $+K_{2z}$. From this it is understood that simultaneous refraction is a two-dimensional effect. In the case of a finite $\sigma$ there will always be oblique waves with $k_x\ne 0$ excited, no matter how large $\sigma$ is. After a sufficiently long time $t$ these oblique waves will somehow establish waves along the $z$-direction with both signs for $K_{2z}$. However, if $\sigma\to\infty$ is taken first, there will be no oblique waves excited. The simultaneous refracting waves can thus not be established. This latter situation is one-dimensional, as the excitation $u_2(x)$ is constant for all $x$, and $K_x=0$.

Trajectories for $k_z$'s branch points, similar to those in Fig. \ref{fig:simrefr}b, can be achieved using a medium with the same refractive index \eqref{refrantiev} as in the previous example, but at a slightly higher observation frequency $\omega_1=0.853\omega_0$. At this frequency, and for sufficiently small $|K_x|$, we have $|\im k_x| < |\re k_x-K_x|$ on the integration path (Fig. \ref{fig:simrefr}c). Then the limit $\sigma\to\infty$ exists, and we end up with the field \eqref{simrefrfieldplanew} for the Fresnel situation, and a similar result for the current source in the plane $z=0$ (then the transmission coefficients $\frac{2\mu_2K_{1z}}{\mu_2K_{1z}\pm\mu_1K_{2z}}$ are replaced by $\pm\mu\omega_1/2K_zc$).

The time domain response of a medium with $\epsilon(\omega)=\mu(\omega)=n(\omega)$, where the refractive index $n(\omega)$ is given by \eqref{refrantiev}, was simulated using the FDTD method \cite{yee66} for Lorentzian media \cite{kelley96}. In the simulation the situation with a current source in $z=0$ was implemented. For $K_x=0$, $\omega_1=0.853\omega_0$, and a finite, but large $\sigma$, the field should describe a partially standing wave consisting of traveling waves with both signs of $K_z$, after sufficiently long time. It turns out, however, that the time it takes to reach the monochromatic limit is much longer than what is possible to simulate. 

The simulations show that the fields grow rapidly as they propagate, both in the $x$ and $z$-direction. This rapid growth is explained as follows. Since the excitation vanishes for $t<0$, it will contain other frequencies than just the observation frequency. Even though the frequency spectrum has a large peak at $\omega_1$, the frequencies around resonance $\omega_0$ will dominate for a very long time, due to extremely high gain there. Indeed, $n(\omega_0)=1-10i$, so at resonance the forward propagating wave will grow as $\exp(20\pi z/\lambda)$, where $\lambda$ is the vacuum wavelength, as it propagates in the $z$-direction. Also the side waves, with $k_x=\pm n(\omega_0)\omega_0/c$, will grow at this rate in the $\pm x$-direction. Since $|\im k_x| > |\re k_x|$ these side waves will be strongly excited. For $t\to\infty$ the excitation only contains $\omega_1$, and the field should eventually describe simultaneous refraction. However, as can be verified using frequency-domain simulations, the transients are extremely strong so it takes a very long time for them to die out.

Due to numerical errors artificial reflections may happen during FDTD if the fields become extremely large. If such artificial reflections occur before the monochromatic limit is reached, the simulation will never be able to reveal simultaneous refraction: waves may be reflected back and forth, being amplified as they propagate, and the solution will eventually grow with time even at fixed points in space.

Nistad and Skaar showed that negative refraction can occur at a single observation frequency $\omega_1$, with arbitrarily low loss for all frequencies, if there is a steep drop in $\im{n(\omega)}$ just below $\omega_1$ \cite{nistad08}. It is similarly possible to achieve a negative refractive index $n=\sqrt{\epsilon\mu}$ at arbitrarily low gain through a steep drop in $\im{n(\omega)}$ just above the observation frequency. For such a medium, the trajectories of $k_z$'s branch points will in fact be similar to those in Fig. \ref{fig:simrefr}b for the frequencies where $n(\omega)<0$. One such medium, where the maximum gain was reduced to $\im{n(\omega)}=-2$, was simulated, but artificial reflections destroy the validity of the simulation solution before the transients die out. For FDTD simulations to be able to reveal simultaneous refraction, media with a significantly lower gain, while having branch point trajectories as in Fig. \ref{fig:simrefr}b, must be found.
 
\section{Discussion and conclusion}\label{sec:concl}
Wave propagation in gain media has been considered by a Fourier--Laplace integral in space and time. How the correct monochromatic and plane wave limits can be taken is demonstrated, by deforming the integration surface in complex frequency-wavenumber space. In some cases it is possible to deform the inverse Laplace transform contour down to the real $\omega$-axis, at the expense of deforming the inverse Fourier $k_x$-integration path. For active media where this can be done, the path will contain complex $k_x$, representing amplified waves as they propagate in the $x$-direction. If such a deformation is not possible, the inverse Laplace transform will contain complex frequencies, and the field will therefore grow exponentially with time, even at a fixed point in space: there is an absolute instability.   

It is shown that the monochromatic and plane wave limits generally do not commute; for example, one order may lead to a diverging field, while the other order leads to a finite field. The plane wave limit may be dependent on whether it is realized by a gaussian excitation or a finite support excitation, eventually of infinite width. This is because amplifying side waves are less excited by the gaussian excitation.

The general path deformation theory is applied to analyze familiar passive and active media, and to predict media with novel properties. In particular it is shown that certain gain media may be simultaneous refracting, i.e. they refract positively and negatively at the same time. It is argued that this is a two-dimensional effect, i.e. it will not occur if an infinitely wide source produces a wave propagating only in the $z$-direction. The monochromatic plane wave response of these media generally depends on which of the limits is taken first, or the width of the source relative to the duration of the experiment as both of these parameters tend to infinity.

An example of a simultaneous refracting medium is given. For a large, but finite width of the source, this medium is, in principle, simultaneous refracting after a sufficiently long time, i.e. in the monochromatic limit. In attempt to visualize the effect, and to independently verify the theory, time domain simulations of this medium were performed. However, the simulations were not able to visualize the effect, as the monochromatic limit never was reached. The suggested medium has a very large gain at resonance, so frequencies of the transients close to resonance will be strongly amplified as they propagate into the medium. Due to the occurrence of artificial reflections before these transients die out, simultaneous refraction is therefore not seen in the simulations. Similar stability problems are expected for experimental realizations. It should therefore be investigated if simultaneous refracting media with significantly less gain exist.

\appendix
\section{Properties of $k_z(k_x,\omega)$}\label{sec:fubini}
We here consider the properties of the function $k_z(k_x,\omega)$ along the real $k_x$-axis, and in a region $\im\omega>\gamma$. The function is defined by \eqref{kz} and \eqref{kzsign}. We prove that $k_z(k_x,\omega)$ is zero-free and analytic in both arguments. Moreover, $k_z\to+\omega/c$ for $\omega\to\infty$ and fixed $k_x$, and $k_z\to+ik_x$ for $k_x\to+\infty$ and fixed $\omega$. Initially, we require $\gamma$ to be large, such that $\epsilon\mu$ is close to unity in the region. In Sec. \ref{sec:cont}, we use analytic continuation to make use of the results in a larger region (i.e., reduce $\gamma$).

\begin{figure}
\centering \includegraphics[width=7.5cm]{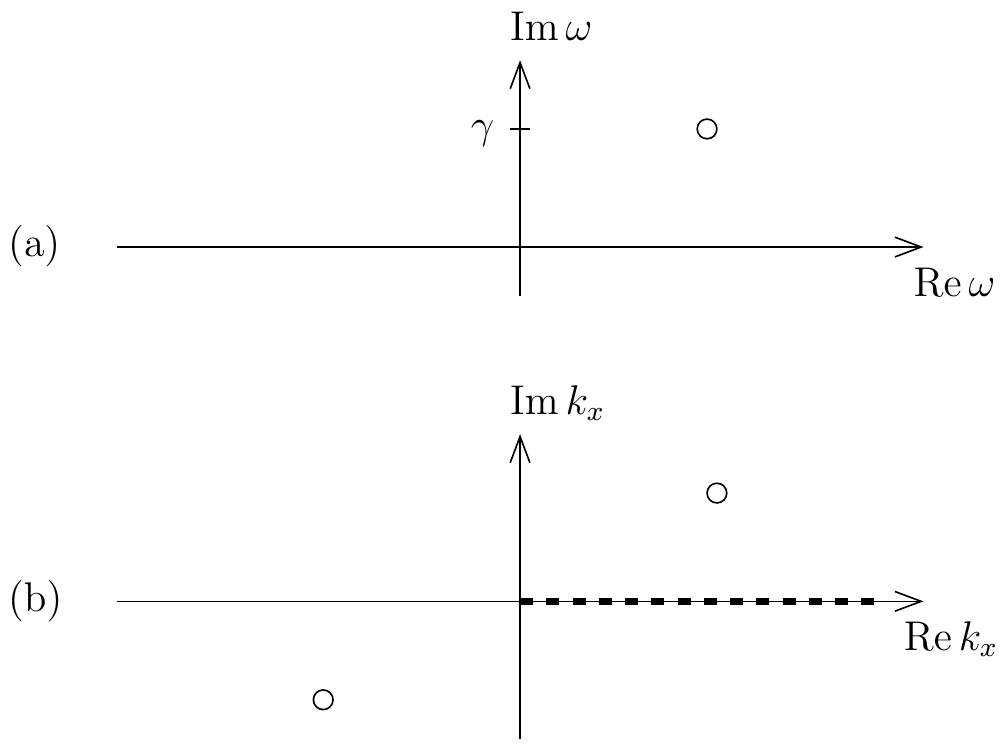}
\caption{For a fixed $\omega$, with $\im\omega>\gamma$ and $\re\omega>0$ (indicated by an open circle in the complex $\omega$-plane (a)), the zeros of $k_z^2=\epsilon\mu\omega^2/c^2-k_x^2$ are shown in the complex $k_x$-plane (b). For large $\im\omega$, the zeros $k_x=\pm\sqrt{\epsilon\mu}\omega/c$ are located away from the real axis.}
\label{fig:kxzeros}
\end{figure}

First, we consider the zeros of $k_z(k_x,\omega)$, given by $k_x=\pm\sqrt{\epsilon\mu}\omega/c$, see Fig. \ref{fig:kxzeros}. None of these is located at real $k_x$, since $\omega$ is complex in the region $\im\omega>\gamma$ and $\epsilon\mu$ is close to unity there: Consider first a region characterized by a bounded $\re\omega$. If a zero existed for positive $k_x$, we could just increase $\gamma$ (and therefore $\im\omega$) such that $\sqrt{\epsilon\mu}$ gets closer to unity and $\arg\omega$ increases; then the zero would move away from the real $k_x$-axis. Next, consider $\re\omega\to\infty$. Since $\sqrt{\epsilon\mu}=1+\mathcal{O}(\omega^{-2})$, the zeros are located at $k_x=\pm\omega/c+\mathcal O(\omega^{-1})$. Thus $k_z(k_x,\omega)$ has no zeros approaching the real $k_x$-axis as $\re\omega\to\infty$.

Second, we argue that $k_z(k_x,\omega)$ is analytic in both arguments. The analyticity in $\omega$ has already been established \eqref{kzsign}, and the analyticity in $k_x$ is immediate from \eqref{kz} provided there are no sign changes. Indeed, such sign changes are impossible: If $k_z(k_x,\omega)$ were discontinuous in $k_x$, we could find a $(k_x,\omega)$ and a tiny $\delta$ such that $k_z(k_x+\delta,\omega)\approx-k_z(k_x,\omega)$. This leads to a contradiction since $k_z(k_x,\omega)$ is zero-free and continuous in $\omega$ in the region $\im\omega>\gamma$, and $k_z(k_x+\delta,\omega)\to k_z(k_x,\omega)$ as $\omega\to\infty$ there.

It is interesting to examine the behavior of $k_z$ in the limit of large $k_x$. The sign of $k_z$ for active media in the total internal reflection regime has been discussed extensively in previous literature \cite{koester66,kolokolov99,grepstad11}. For $k_x=0$, we have $k_{z}\approx\omega/c$ in the region $\im\omega>\gamma$. As $k_x$ increases along the dashed line in Fig. \ref{fig:kxzeros}, the complex argument of $k_z^2$ increases according to the zero configuration in the figure. Since $k_z$ is a continuous function of $k_x$ it follows that as $k_x\to+\infty$, $k_z\to+ik_x$. This seems to predict an evanescent behavior in the total internal reflection regime of large $k_x$; however, it is important to remember that we only have considered the complex frequencies with $\im\omega>\gamma$. Interpretation at real frequencies is possible under certain circumstances (Section \ref{sec:weakgain}) \cite{kolokolov99,grepstad11}; however, for conventional, weak gain media it turns out to be an instability associated with amplified side waves.

\section{Existence of transforms and interchanging the order of integration}\label{sec:fubini2}
Here we establish the existence of the involved transforms in solving Maxwell's equations, and argue that their order can be interchanged. To establish the existence, we must make assumptions on the electric and magnetic fields, and their derivatives wrt. $x$, $z$, and $t$. These assumptions enable formulating electromagnetics in the $(k_x,\omega)$-domain by the $\mathcal L^2$ theory of Fourier transforms, to obtain the solutions \eqref{Ezpos}. Finally, we verify that the solutions are consistent with the initial assumptions, making a self-consistent theory.

To limit the amount of writing, we will only consider the electric field $\mathcal E(x,z,t)$ here; the other functions can be treated similarly with some small complications from derivatives. We will only consider the solution \eqref{Ezposa}; the other solution \eqref{Ezposb} can be treated similarly. With respect to $x$ the function $\mathcal E(x,z,t)$ is assumed to be in the Hilbert space $\mathcal L^2$ of square integrable functions. With respect to $t$, $\mathcal E(x,z,t)\exp(-\gamma t)$ is assumed to be in $\mathcal L^2$, for a sufficiently large, positive $\gamma$. Defining
\begin{subequations}\label{1transf}
\begin{align}
& E(x,z,\omega) = \int_{0}^{\infty} \mathcal E(x,z,t)\exp(i\omega t)\diff t, \\
& E(k_x,z,t) = \int_{-\infty}^{\infty} \mathcal E(x,z,t)\exp(-ik_xx)\diff x,
\end{align}
\end{subequations}
we assume that $E(x,z,\omega)$ is in $\mathcal L^2$ wrt. $x$ for $\im\omega=\gamma$, and $E(k_x,z,t)\exp(-\gamma t)$ is in $\mathcal L^2$ wrt. $t$ for real $k_x$. This means that we can Fourier transform $E(x,z,\omega)$ wrt. $x$, or Laplace transform $E(k_x,z,t)$. By solving Maxwell's equation in the resulting transform domain $(\omega,k_x)$ we obtain \eqref{Ezposa}. Our job now is to verify all assumptions, after inverse transformation of \eqref{Ezposa}.

To this end, we assume that the source $u(x)v(t)$ is sufficiently smooth such that
\begin{subequations}\label{sourcecond}
\begin{align}
& U(k_x)k_x^p \in \mathcal L^1 \cap \mathcal L^2, \\
& V(\omega)\omega^p \in \mathcal L^1 \cap \mathcal L^2, 
\end{align}
\end{subequations}
for $p=0$, 1, and 2. That a function of $\omega$ is in $\mathcal L^1\cap\mathcal L^2$, such as e.g. $V(\omega)$, is to be interpreted as $V(\omega'+i\gamma) \in \mathcal L^1\cap\mathcal L^2$ viewed as a function of the real variable $\omega'$. 

Consider the factor $\mu\exp(ik_zz)/k_z$ in \eqref{Ezposa}. A little thought shows that this factor is bounded along the integration surface $(-\infty,\infty)\times (i\gamma-\infty,i\gamma+\infty)$ in the $(k_x,\omega)$-space. Thus $E(k_x,z,\omega) \in \mathcal L^1$ wrt. $(k_x,\omega)$, so with the help of Fubini's theorem we can express $\mathcal E(x,z,t)$ with inverse transforms of either order \eqref{invlaplfour}.

By taking only one of the inverse transforms in \eqref{invlaplfour}, we can write
\begin{subequations}\label{1inv}
\begin{align}
& E(x,z,\omega) = \frac{1}{2\pi}\int_{-\infty}^{\infty} E(k_x,z,\omega)\e{ik_xx}\diff k_x, \\
& E(k_x,z,t) = \frac{1}{2\pi}\int_{i\gamma-\infty}^{i\gamma+\infty} E(k_x,z,\omega)\e{-i\omega t}\diff\omega.
\end{align}
\end{subequations}
Clearly, $E(k_x,z,\omega)\in\mathcal L^2$ both wrt. $k_x$ and $\omega$, so the functions $E(x,z,\omega)$ and $E(k_x,z,t)\exp(-\gamma t)$ is in $\mathcal L^2$ wrt. $x$ and $t$, respectively. 

Substituting \eqref{Ezposa} into \eqref{1inv} we obtain
\begin{subequations}\label{1invsubst}
\begin{align}
& E(x,z,\omega) = \frac{\mu V(\omega)\omega}{4\pi c}\int_{-\infty}^{\infty} \frac{\e{ik_zz}}{k_z}U(k_x)\e{ik_xx}\diff k_x, \label{1invsubsta}\\
& E(k_x,z,t) = \frac{U(k_x)}{4\pi c}\int_{i\gamma-\infty}^{i\gamma+\infty} \frac{\mu\e{ik_zz}}{k_z}V(\omega)\omega\e{-i\omega t}\diff\omega. \label{1invsubstb}
\end{align}
\end{subequations}
Since the integrals in \eqref{1invsubsta} and \eqref{1invsubstb} are bounded wrt. $\omega$ and $k_x$, respectively, it follows that $E(x,z,\omega)$ and $E(k_x,z,t)$ are in $\mathcal L^2$ wrt. $\omega$ and $k_x$. After the final inverse transforms we therefore obtain a function $\mathcal E(x,z,t)$ for which $\mathcal E(x,z,t)\exp(-\gamma t)\in \mathcal L^2$ wrt. $t$ and $x$.

We have assumed that the excitation $u(x)v(t)$ is sufficiently smooth, such that $U(k_x)$ and $V(\omega)$ tend to zero sufficiently quickly \eqref{sourcecond}. Considering the source $U_2(k_x)$, as given by \eqref{Usources}, this is automatically satisfied. For $U_1(k_x)$ the condition will be satisfied if $u_1(x)$ is three times differentiable at its endpoints, as evident from \eqref{asymptU1}. For $V(\omega)$, as given by \eqref{laplexcit}, the condition will not be satisfied; however this can be fixed by slightly smoothening the onset of $v(t)$ such that it is three times differentiable. This makes \eqref{laplexcit} valid for arbitrarily large $\omega$, before it starts to decay faster. Defining $\tilde v(t)$ as the smoothened excitation, $\tilde v(t)-v(t)$ has finite support. Thus $\tilde V(\omega)-V(\omega)$ is an entire function, which means that the smoothening does not affect the electric field solution in the monochromatic limit.

The reason for doing this analysis in a more rigorous way than is common in the physics literature, is the appearance of unusual phenomena and the need to go back to first principles when considering gain media. Nevertheless, this appendix shows that the conditions for existence of the transforms are similar for active and passive media; the only difference is that the Fourier transform in time (for passive media) must be replaced with the Laplace transform (for active media).
 
\section{Causality and the Titchmarsh theorem for diverging functions}\label{sec:Titchdiv}
To prove the causality result \eqref{determABCDfin} from the choice \eqref{kzsign}, we employ the Titchmarsh theorem \cite{nussenzveig, titchmarsh}, formulated for exponentially diverging functions. 

Let $f(t)$ be a causal function,
\begin{equation}\label{causalE}
f(t)=0, \quad\text{for } t<0,
\end{equation}
such that 
$f(t)\exp(-\gamma t)$ is square integrable for some real $\gamma$. Consider the Laplace transform of $f(t)$:
\begin{equation}
F(\omega)=\int_0^\infty f(t)\exp(i\omega t)\diff t.
\end{equation}
Then
\begin{equation}\label{anal}
F(\omega) \text{ is analytic for } \im\omega>\gamma, 
\end{equation}
and there is a uniform bound $K$ such that
\begin{equation}\label{unbound}
\int_{-\infty}^{\infty}|F(\omega'+i\gamma')|^2\diff\omega' \leq K <\infty, \quad\text{for all $\gamma'\geq\gamma$.}
\end{equation}
The converse result is also true: Let a function $F(\omega)$ be analytic for $\im\omega>\gamma$ and satisfy \eqref{unbound} for some $K$. Then, the inverse Laplace transform 
\begin{equation}\label{invlapl}
f(t)=\frac{1}{2\pi}\int_{i\gamma-\infty}^{i\gamma+\infty} F(\omega)\exp(-i\omega t)\diff\omega
\end{equation}
satisfies \eqref{causalE} and $f(t)\exp(-\gamma t)$ is square integrable. The proof is immediate from the Titchmarsh theorem by considering the function $f(t)\exp(-\gamma t)$ and its Laplace (or Fourier) transform $F(\omega'+i\gamma)$.

Returning to the electric and magnetic fields for $z>0$, as expressed by \eqref{EHm}, we know that the corresponding time-domain fields are causal \eqref{causalE}. Thus the fields satisfy \eqref{anal} and \eqref{unbound}. This means that $C(k_x,\omega)\e{ik_zz}$ and $D(k_x,\omega)\e{-ik_zz}$, separately, satisfy these conditions. From the initial value theorem and the fact that $k_zc=\omega+\mathcal O(\omega^{-1})$, it is intuitively clear that the factor $\e{-ik_zz}$ shifts the beginning of the associated time-domain response to earlier times by an amount $z/c$. Thus $D(k_x,\omega)\e{-ik_zz}$ can only be compatible with causality for all $z>0$ if $D(k_x,\omega)\equiv 0$. A rigorous argument goes as follows (here we suppress the $k_x$-dependence for clarity): Since $D(\omega)\exp(-ik_zz)$ is required to satisfy \eqref{unbound} for all $z$, and since $k_zc=\omega+\mathcal O(\omega^{-1})$, we have for sufficiently large $\gamma$:
\begin{equation}\label{Dineq}
\int_{-\infty}^{\infty}|D(\omega'+i\gamma)|^2\diff\omega' \leq 2K(z)\exp(-2\gamma z/c).
\end{equation}
Here $K(z)$ is independent of $\gamma$. If $d(t)$ is the inverse Laplace transform of $D(\omega)$, then $d(t)\exp(-\gamma t)$ is the inverse Laplace transform of $D(\omega+i\gamma)$. Thus, from Parseval's relation, $\int_0^\infty |d(t)|^2\exp(-2\gamma t)\diff t=\frac{1}{2\pi}\int_{-\infty}^{\infty}|D(\omega'+i\gamma)|^2\diff\omega'$. Combination with \eqref{Dineq} yields
\begin{align}
 & \int_{0}^{T}|d(t)|^2\diff t \leq \exp(2\gamma T)\int_{0}^{T}|d(t)|^2\exp(-2\gamma t)\diff t  \nonumber\\
 & \leq \exp(2\gamma T)  \int_{0}^{\infty}|d(t)|^2\exp(-2\gamma t)\diff t \\
 &\leq \frac{K(z)}{\pi}\exp(-2\gamma (z/c-T)). \nonumber
\end{align}
valid for any $z$ and $T$, and for sufficiently large $\gamma$. Letting $z/c>T$, it is apparent that we can make the right-hand side as small as we wish, by letting $\gamma$ be sufficiently large. Since $T$ was arbitrary, $d(t)$ vanishes almost everywhere.

\def\cprime{$'$}
%


\end{document}